\documentclass[12pt,preprint]{aastex}
\usepackage{emulateapj5,apjfonts}
\usepackage{graphics}
\usepackage{amssymb}
\usepackage{bbm}
\usepackage{amsmath,textcomp,array}
\usepackage{onecolfloat}
\usepackage[ruled]{algorithm}
\usepackage{color}
\usepackage[normalem]{ulem}
\usepackage{bm}

\lefthead{Lawrence et al.}
\righthead{Mira-Titan Universe II}

\begin{document}


\def\head{

\title{The Mira-Titan Universe II: Matter Power Spectrum Emulation}
\author{
Earl Lawrence\altaffilmark{1}, Katrin~Heitmann\altaffilmark{2,3},  Juliana~Kwan\altaffilmark{4,5}, 
Amol Upadhye\altaffilmark{6},
Derek Bingham\altaffilmark{7},
Salman~Habib\altaffilmark{2,3}, David Higdon\altaffilmark{8}, 
Adrian Pope\altaffilmark{9}, 
Hal Finkel\altaffilmark{9}, Nicholas Frontiere\altaffilmark{2,10}}

\affil{$^1$ CCS-6, CCS Division, Los Alamos National Laboratory, Los Alamos, NM 87545, USA}
\affil{$^2$ HEP Division,  Argonne National Laboratory, Lemont, IL 60439, USA}
\affil{$^3$ MCS Division, Argonne National Laboratory, Lemont, IL 60439, USA}
\affil{$^4$ Kavli Institute for the Physics and Mathematics of the Universe (Kavli IPMU, WPI), Todai Institutes for Advanced Study,
the University of Tokyo, Kashiwa, Chiba 277-8583, Japan}
\affil{$^5$ Department of Physics and Astronomy, University of Pennsylvania, Philadelphia, PA 19104, USA}
\affil{$^6$ Department of Physics, University of Wisconsin-Madison, Madison, WI 53706, USA}
\affil{$^7$ Department of Statistics and Actuarial Science, Simon Fraser University, Bunraby, BC, Canada }
\affil{$^8$ Social and Decision Analytics Laboratory, Virginia Bioinformatics Institute, Virgina Tech, Arlington, VA 22203, USA}
\affil{$^9$ ALCF Division,  Argonne National Laboratory, Lemont, IL 60439, USA}
\affil{$^{10}$ Department of Physics, University of Chicago, Chicago, IL 60637, USA}

\date{today}

\begin{abstract} 

We introduce a new cosmic emulator for the matter power spectrum
covering eight cosmological parameters. Targeted at optical surveys,
the emulator provides accurate predictions out to a wavenumber $k\sim
5$Mpc$^{-1}$ and redshift $z\le 2$. Besides covering the standard set
of $\Lambda$CDM parameters, massive neutrinos and a dynamical dark
energy of state are included. The emulator is built on a sample set of
36 cosmological models, carefully chosen to provide accurate
predictions over the wide and large parameter space. For each model,
we have performed a high-resolution simulation, augmented with sixteen
medium-resolution simulations and TimeRG perturbation theory results
to provide accurate coverage of a wide $k$-range; the dataset
generated as part of this project is more than 1.2Pbyte. With the current
set of simulated models, we achieve an accuracy of approximately 4\%.
Because the sampling approach used here has established convergence
and error-control properties, follow-on results with more than a
hundred cosmological models will soon achieve $\sim 1\%$ accuracy. We
compare our approach with other prediction schemes that are based on
halo model ideas and remapping approaches. The new emulator code is publicly available.  

\end{abstract}

\keywords{methods: statistical ---
          cosmology: large-scale structure of the universe}}

\twocolumn[\head]

\section{Introduction}

The field of cosmology has undergone a remarkable transformation in
the last two decades -- from a somewhat qualitative picture of the
make-up and evolution of the Universe we have arrived at the `Standard
Model' of cosmology with parameters constrained at the few percent
level~\citep{ade15, anderson14}.  Despite the phenomenological and
descriptive success of the Standard Model, many foundational questions still
require answers. Our understanding of the fundamental physics is
lacking in critical areas: We do not properly understand the cause of
the accelerated expansion of the Universe~\citep{caldwell09}, the
nature of dark matter is unknown~\citep{feng10}, and our understanding
of the physics of inflation remains incomplete, to mention three of
the most prominent puzzles. Ongoing and upcoming cosmological surveys
and experiments aim to address these and other questions by providing
datasets with much smaller statistical errors and extended range in
spatial scales and redshift. Analysis of these observations can
provide important clues by providing evidence against a cosmological
constant, or even against general relativity as the preferred theory
of gravity~\citep{joyce15}. Some of the data will put added
constraints on the properties of dark matter candidates, and also
significantly tighten the current cosmological errors on determining
the sum of neutrino masses. Because of the enhancement of data
quality, it is therefore important -- from a theoretical and modeling
perspective -- to open up new parameters beyond the standard set of
$\theta=\{\omega_{cdm}, \omega_b,\sigma_8, h, n_s\}$ and also enter
new uncharted areas with respect to length scales, exploring nonlinear
regimes that might provide new insights into the dynamics of the
Universe.

In order to take full advantage of the new data, and not be
theory/modeling-limited, prediction tools must be available at
accuracy levels significantly better than those characteristic of the
measurements. Since surveys increasingly probe the nonlinear
regime of structure formation, theoretical predictions have to be
derived from detailed and error-controlled simulations that are
necessarily computationally expensive. While this may be a reasonable
approach to study individual models, it is not practical as a tool for
exploring parameter space, nor does it help in solving the
cosmological inverse problem of determining parameters based on
observational knowledge of a set of summary statistics, where hundreds
of thousands to millions of forward model evaluations may be needed.

In order to address the above requirement, we have embarked on a
program to create very fast oracles, or ``cosmic emulators'' for
various cosmic probes.  The aim of the approach is to achieve robustly
accurate prediction schemes over a range of cosmological parameters,
based on a relatively small number of underlying simulations. The
complete framework not only provides predictions for specific
cosmological statistics but also includes a self-contained Bayesian
inference engine to constrain cosmological parameters by combining
observational data and emulator predictions (`cosmic calibration'). We
first introduced the concept in~\cite{HHHN} based on a set of lower
resolution gravity-only simulations and a simulated data set for the
nonlinear matter power spectrum. In a second paper, \cite{HHHNW}, we
extended the approach to include measurements for the cosmic microwave
background (CMB). In a following set of four
papers~\citep{coyote1,coyote2,coyote3,emu_ext}, the Coyote Universe
series, we focused on providing a high-accuracy prediction tool for
the matter power spectrum over a range of six cosmological parameters,
adding $w$ to the standard set of five. Later, we added other
emulators to predict the halo concentration-mass
relation~\citep{kwan13} and the galaxy power spectrum, using halo
occupation distribution (HOD) modeling~\citep{kwan14}.

In this paper, we focus again on the matter power spectrum, while
extending the range of cosmological parameters to eight,
$\theta=\{\omega_{cdm}, \omega_b,\sigma_8, h, n_s, w_0, w_a,
\omega_\nu\}$ and employing higher-quality simulations than in the
original Coyote Universe, simultaneously improving on the mass
resolution and the simulation volume. This work builds on the
convergent sampling strategy described in~\cite{heitmann15} to
systematically improve emulation accuracy by adding new simulations to
a previous sample, following \cite{Bergner:2011}. In the original
work, the idea was demonstrated on a set of linear power spectra as
well as for mass function predictions (assuming universality of the
mass function across cosmologies, which is valid at the 5-10\% level)
-- we now release the first nonlinear emulator from the Mira-Titan
Universe simulation suite. The inclusion of a dynamical dark energy
component and massive neutrinos is nontrivial -- the approach to the
simulations is described in more detail in~\cite{heitmann15}. Several
tests of the simulation methodology were carried out in~\cite{upadhye14},
where results on large scales were compared to TimeRG perturbation theory. Discussions of the range of validity and methods for adding baryonic corrections are provided in Section~\ref{subsec:validity}.

The eventual aim of the emulators constructed from the Mira-Titan
Universe simulation suite is to reach simulation prediction accuracies
at the $1\%$ level, which requires results for more than a hundred
cosmological models. The sampling strategy followed allows us to make
emulators at intermediate accuracy levels before all the simulations
are completed and to check thereby that the appropriate accuracies are
in fact being achieved during this process. (This also includes
demonstrating successful data filtering, data reduction with principal
components, and finally, Gaussian process modeling to carry out the
required interpolation.) Results presented in this paper demonstrate
the success of this strategy.

Several other approaches have been suggested to provide predictions
for the matter power spectrum going beyond $\Lambda$CDM.
\cite{agarwal14} included neutrinos and constructed an emulator using
machine learning techniques. \cite{takahashi12} used a set of
simulations to improve the original Halofit predictions
and~\cite{bird} added a neutrino contribution to this model.
\cite{casarini} used an approximate approach to extend the Coyote
Universe emulator to include $w_a$ as a new parameter, in order to
cover the model space of dynamical dark energy models. Finally,
\cite{mead16} used a halo model approach to include effects of
neutrinos, modified gravity, and dynamical dark energy. We will
discuss these approaches and compare some of their results with the
emulator presented here in Section~\ref{comparison}.
will
The paper is organized as follows. In Section~\ref{params} we describe
the cosmological parameter space covered and provide relevant details
of the simulation suite used to build the emulator. In
Section~\ref{emu} we discuss emulator construction with a focus on
error estimates. We compare our results to other approaches in
Section~\ref{comparison}, ending with a summary and outlook in
Section~\ref{conclusion}. The emulator is publicly available via a github repository\footnote{https://github.com/lanl/CosmicEmu} and on our CosmicEmu webpage\footnote{http://www.hep.anl.gov/cosmology/CosmicEmu/emu.html}.

\section{Parameter Ranges and Simulations}
\label{params}
The parameter range now allows for dynamic dark energy and varying the neutrino mass sum. The Mira-Titan Universe suite of simulations is well on its way to completion; we describe below the general characteristics of the simulations, including a separate discussion of how neutrinos are included.

\subsection{Parameters}
The choices for the parameter ranges covered in this paper are
discussed in detail in~\cite{heitmann15}. In addition to the five
standard parameters describing the $\Lambda$CDM model, we include a
dynamical dark  energy equation of state, parameterized by
$(w_0,w_a)$, and massive neutrinos. We fix the effective number of
neutrino species to be $N_{eff}=3.04$. The dark energy equation of state is
parameterized in the standard form:
$w(a)=w_0+w_a(1-a)$~\citep{chevalier,linder}. Our parameter ranges are 
informed by recent observations of the CMB and large scale optical
surveys. In addition, we aim to cover the relevant ranges for ongoing
and upcoming surveys. With these considerations in mind, we choose the
following ranges over the eight cosmological parameters (with a flat prior assumption): 
\begin{eqnarray}\label{cosmoparams}
0.12\le &\omega_m& \le 0.155,\\
0.0215\le &\omega_b& \le 0.0235,\\
0.7\le &\sigma_8& \le 0.9,\\
0.55\le &h& \le 0.85,\\
0.85\le &n_s& \le 1.05,\\
-1.3\le &w_0& \le -0.7,\\
-1.73\le &w_a& \le 1.28,\\
0.0\le &\omega_\nu& \le 0.01.
\end{eqnarray}
Note that $w_a$ is actually jointly constrained with $w_0$ such that $0.3 \leq (-w_0-w_a)^{1/4}$. See~\cite{heitmann15} for a discussion.

\subsection{Simulations}
The large-scale simulations described in this paper were carried out
with the HACC (Hardware/Hybrid Cosmology Code) framework, a
high-performance cosmology code, designed to take advantage of current
and future supercomputer architectures; HACC is described in detail
in~\cite{habib16}. HACC simulations were carried out on the Mira
supercomputer at the Argonne Leadership Computing Facility (ALCF) and
on the Titan system at the Oak Ridge Leadership Computing Facility
(OLCF). Mira belongs to the family of IBM's Blue Gene Q systems (BG/Q)
and has 786,432 compute cores, while Titan achieves its high
performance due to the NVIDIA K20 Graphics Processing Units (GPUs)
attached to each of its $\sim 18,000$ compute nodes. The
simulations described in this paper are modest in size compared to the
capabilities of these machines, but the sheer number of simulations
needed for our full program ($\sim 100$) makes this work
computationally expensive. We note that HACC uses different algorithms on the above systems but the results for the power spectrum agree to within small fractions of a percent~\citep{habib16}, much smaller than the final target error of the emulator.

The results for each sampled cosmological model were obtained as
follows. We first evaluate the power spectrum using the TimeRG
perturbative approach (introduced in~\citealt{timeRG1}), as described
in~\cite{upadhye14}. This provides a smooth and very accurate
prediction of the power spectrum on large scales (small $k$), out to
$k\sim0.04$Mpc$^{-1}$ for $0 \le z \le 1$ and $k\sim 0.14$Mpc$^{-1}$
for $z\le 2$. Several N-body simulations were carried out next. In
order to cover the intermediate scales (out to $k\sim
0.25$Mpc$^{-1}$), we use 16 realizations of particle mesh (PM)
simulations carried out with HACC. These simulations evolve 512$^3$
particles on a 1024$^3$ grid and cover a volume of (1300Mpc)$^3$
each. For the small scale (high $k$) regime we carry out one
high-resolution simulation with HACC per cosmology. These simulations
evolve 3200$^3$ particles starting at $z_{in}=200$ using the
Zel'dovich approximation, each in a (2100Mpc)$^3$ volume, leading to
a mass resolution of approximately $\sim$10$^{10}$M$_\odot$, depending
on the detailed cosmological parameters. The force resolution of these
simulations is $\sim 6.6$kpc. For each of the high resolution runs we
store a range of outputs:

\begin{itemize}
\item Particle outputs (full and randomly down-sampled to 1\%) at the
  following redshifts: $z=\{$4.00, 3.04, 2.48, 2.02, 1.78, 1.61, 1.38,
  1.21, 1.01, 0.78, 0.74, 0.70, 0.66, 0.62, 0.58, 0.54, 0.50, 0.47,
  0.43, 0.40, 0.36, 0.30, 0.24, 0.21, 0.15, 0.10, 0.0$\}$ 
\item Halo information at the same redshifts for friends-of-friends
  halos with a linking length of $b=0.168$ with at least 20 particles
  per halo, halo centers are based on a potential minimum evaluation
\item Halo information at eight redshifts for friends-of-friends halos
  with a linking length of $b=0.2$ with at least 20 particles per
  halo; halo centers are based on a potential minimum evaluation
\item Halo information at the same redshifts for spherical overdensity
  halos with $M_{200}$ with at least 1,000 particles per halo
\item Halo information at eight redshifts for spherical overdensity
  halos with $M_{300}$, $M_{500}$ with at least 1,000 particles per halo
\item All particles that reside in halos with at least 1,000
  particles, 1\% of particles in smaller halos, randomly selected, at
  least 5 particles per halo 
\item Particle and halo tags for all particles in halos
\item Power spectra at the same redshifts, though only the following
  are used for building the emulator: $z=\{$2.02, 1.61, 1.01, 0.66,
  0.43, 0.24, 0.10, 0.0$\}$ 
\end{itemize}
Keeping this data leads to an uncompressed dataset size of
approximately 38TB per model, and more than 1PB for the simulation
suite discussed in this paper. Storing the relatively large number of
time slices allows for creating light-cones from the outputs,
following the approach presented in~\cite{sunayama16}. For the power
spectrum emulator, generation of a sub-set of the power spectrum
measurements is sufficient due to the smooth evolution of $P(k)$. In
addition, many more emulators can be created from this data set, for
quantities such as the mass function, galaxy correlation function,
etc. While it is difficult to make the full dataset publicly available
(the raw particle outputs will reside on tape for long-term storage
and retrieval is currently slow), we are planning to make the
processed data, such as the halo catalogs, publicly available in the
near future.

\subsubsection{Treatment of Neutrinos}
\label{sec:neutrinos}

The treatment of neutrino effects in cosmological simulations is
nontrivial. This is mainly due to two issues: 1) the very high
neutrino thermal velocities early on in the simulation, and 2) the
very large mass ratio between the dark matter tracer particles and the
neutrino tracer particles. Many solutions to these problems have been
discussed in the literature, from adding the neutrinos only at late
times, when the thermal velocities are much smaller (helping with the
first but not the second problem), to introducing coarser force
resolution for the neutrinos to avoid the second problem, to treating
the neutrinos perturbatively (for more details, see, e.g.,
\cite{agarwal,bird,brandbyge1,brandbyge2,brandbyge3,gardini,
inman15,klypin,viel,banerjee} and references therein).

We follow the approach discussed in detail in~\cite{heitmann15},
applying a small correction to account for the scale dependence of the
growth function as discussed in~\cite{upadhye14}. We provide a short
summary of our neutrino treatment here and show comparisons to results
found by other groups in Section~\ref{comparison}. Since we consider
the case of relatively small neutrino masses, the most conservative
treatment of neutrinos suffices. In this treatment, the neutrinos are
not evolved as a separate species, but the linearly evolved neutrino
component is added at each redshift separately. At $z=0$, the
simulation is normalized to the full linear neutrino-baryon-CDM power
spectrum as given by CAMB~\citep{camb}. The baryon-CDM component is
taken to the starting redshift with a scale-independent growth
function and evolved forward with the N-body code, including the
neutrino component in the background equations. This is done for
consistency, since the forward evolution does not have the
scale-dependent growth characteristic of massive neutrinos. At each
redshift of interest, we add the linear neutrino power spectrum to the
nonlinear baryon-CDM component. This approach is valid as long as the
neutrino density fraction $f_{\nu}\equiv\Omega_{\nu}/\Omega_m$ is
sufficiently small.

The result of the procedure outlined above is a low-redshift power
spectrum that accurately includes nonlinearity in the CDM $+$ baryon
sector as well as neutrinos treated linearly. \cite{castorina15} have
found this assumption to be accurate at the $1\%$ level for neutrino
masses satisfying current bounds, when compared with N-body
simulations that include massive neutrinos as separate particles.
Meanwhile, at higher redshifts $z \gtrsim 1$, our use of the
scale-independent CDM $+$ baryon growth factor leads to an error at
large scales where neutrinos cannot be neglected. Fortunately, these
scales are linear, and \cite{upadhye14} showed that the resulting
error can be removed by multiplying the N-body power spectrum by the
$k$-dependent correction factor $D_{\mathrm{b+CDM+}\nu}(k,z)^2 /
D_\mathrm{b+CDM}(z)^2$. Here $D_{\mathrm{b+CDM+}\nu}$ and
$D_\mathrm{b+CDM}$ are, respectively, the linear growth factors for
baryons + CDM + $\nu$ and baryons + CDM. The corrected N-body power
spectrum is consistent with perturbation theory at large scales to
within the simulation error bars. This procedure can be interpreted in the separate universe sense where it has been shown that the scale-dependent clustering of an extra field (quintessence or neutrinos) can be neglected (up to some level of approximation) when the simulation box size is smaller than the Jeans length of that field~\citep{hu16, chiang16}. Our neutrino treatment is reasonable below the Jeans (neutrino free-streaming) scale, and we correct it on super-Jeans-scales. 

\section{Emulator Construction and Testing}
\label{emu}
In this Section we discuss the range and validity of the emulator including the possibility of adding baryonic corrections in post-processing, the smoothing procedure applied to the power spectrum from the set of individual simulations to produce the mean spectrum, and the final process of constructing the emulator. In the latter two cases we discuss the associated errors and how they are estimated and checked.

\subsection{Range of Validity}
\label{subsec:validity}
Upcoming surveys require accurate predictions of the matter power
spectrum at levels of a fraction of a percent. The error budget is
complicated as many interacting sources of uncertainty are present.
First, the numerical accuracy of the underlying cosmological
simulations induces an irreducible error. We follow here the
discussions in~\cite{coyote1} in setting the starting redshift and
force and mass resolution. Given our simulation specifications, the
numerical accuracy at the redshifts ($z\le 2$) and scales ($k\le
5$Mpc$^{-1}$) of interest is at the percent level. Second, there are
errors due to the emulation scheme; we discuss these and estimate
their values below. The main source for inaccuracy is the limited
number of models that we consider here. As stated earlier, this error
will reduce significantly as more models are added. Finally, the
largest systematic uncertainty is due to the modeling of neutrinos and
incomplete knowledge of baryonic effects.

Strictly quantifying the error of the neutrino treatment is difficult
since no error-controlled, fully self-consistent, neutrino simulation
exists currently. As explained in Section~\ref{sec:neutrinos}, we
treat the neutrinos not as a separate species but evolve them only in
the background, and we add the linear neutrino power spectrum to
the $P_{cb}$-component obtained from the simulation rather than
compute the nonlinear neutrino power spectrum. The validity of the
first assumption was tested in~\cite{upadhye14} using TimeRG
perturbation theory. In the regime that the perturbative approach is
valid, the agreement was excellent. The second assumption,
investigated in detail in \cite{castorina15}, holds at the 1\% level.

A similar situation holds for the uncertainty due to the lack of a
baryonic treatment. In our simulation, the baryons are only included
in the initial transfer function and gas dynamics and star formation
and feedback effects are not modeled. Baryonic effects on the power
spectrum remain inconclusive due to uncertainties in the modeling of
several effects, such as feedback from active galactic nuclei (AGN) and supernovae (SNe). Attaining predictive
control at the percent level at smaller length scales ($k>
1$Mpc$^{-1}$) is difficult due to these uncertainties.

An alternative approach to carrying out a large number of expensive
hydrodynamics simulations was put forward in~\cite{mead15}. Here the
authors incorporate baryonic effects into a halo model approach and
are able to reproduce results from full hydrodynamics simulations at
the 5\% level of accuracy. In the same spirit, one could model
baryonic effects given emulator predictions if reliable results from
hydrodynamics simulations are available. \cite{zentner} follow a
similar path targeting the convergence power spectrum in modeling
baryonic effects by varying the halo concentration. Approaches where
the baryonic physics is modeled on top of the matter power spectrum
informed by a small number of hydrodynamics simulations will be the
only viable option for the foreseeable future. \cite{eifler} propose
using a PCA decomposition of the OWLs suite of simulations to
parameterize the effect of baryons on the matter power spectrum, which
can then be included in cosmological model fitting and subsequently
marginalized. \cite{kitching} and \cite{maccrann} introduce another
method to account for the impact of baryons in the dark matter power
spectrum by multiplying the Halofit power spectrum by the ratio of the
OWLs dark matter and baryonic power spectrum over the OWLs dark only
power spectrum. Typically the most extreme OWLs simulation is chosen,
the AGN feedback scenario, such that the impact of baryonic effects is
maximized to provide an upper bound. This technique was applied by
the~\cite{DES_cosmic_shear} and~\cite{kwan17} to estimate the effect
of baryonic physics on the power spectrum of cosmic shear and
tangential shear for galaxy-galaxy lensing respectively.

With increasing computing power, a better understanding of
uncertainties in sub-grid modeling, and more observational data for
cross-calibration, the situation will likely improve over time. Given
current uncertainties, it is nevertheless difficult to state an absolute
error on the full matter power spectrum over the range of scales
considered in this paper.

\subsection{Smoothing}
Our overall approach uses ideas described in the Coyote Universe
series of papers~\citep{coyote1,coyote2,coyote3,emu_ext}. In this
subsection and the next, we briefly describe the approach, referring
the reader to the earlier work for further details.

\begin{figure}[h]
\includegraphics[width=3.5in]{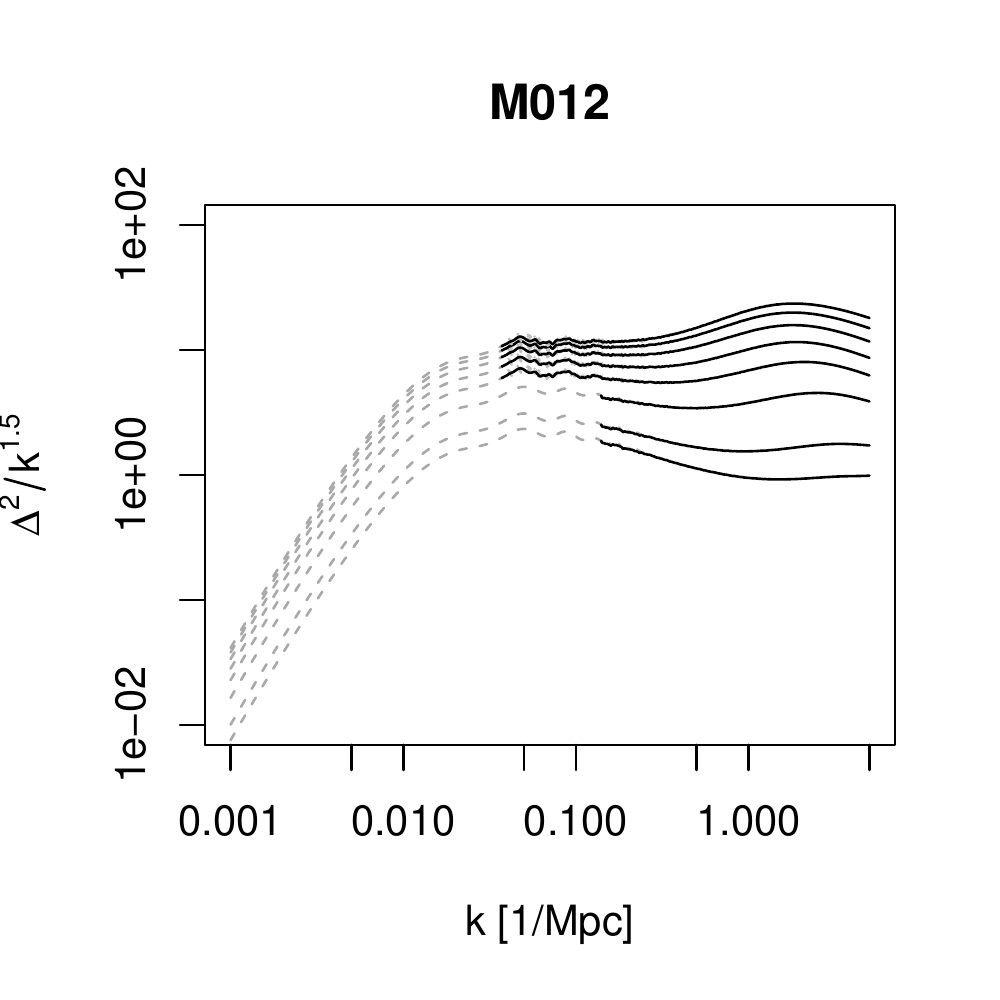}
\caption{Data from the M012 run used in the process
  convolution estimation of the smooth  spectra. The gray dashed lines
  show TimeRG perturbation theory results used at low $k$. The dotted
  gray lines show the low resolutions runs at medium $k$. The solid
  black line shows the high resolution run at medium and high $k$. In
  ascending order, these are at redshifts 2.020, 1.610, 1.006, 0.656,
  0.434, 0.242, 0.101, and 0.000.}
\label{fig:M012raw}
\end{figure}

\begin{figure}
\includegraphics[width=3.5in]{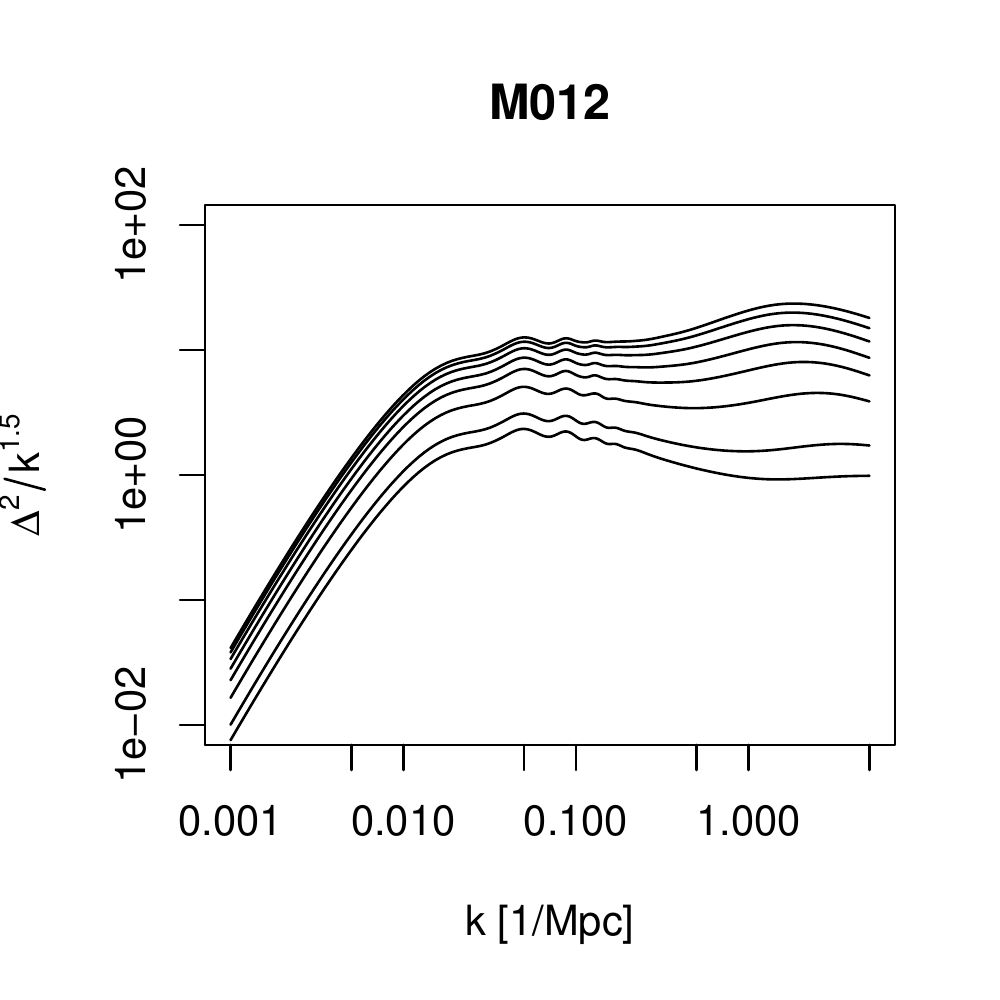}
\caption{Smooth power spectra estimates for the M012 results. In
  ascending order, these are at redshifts 2.020, 1.610, 1.006, 0.656,
  0.434, 0.242, 0.101, and 0.000.} 
\label{fig:M012smooth}
\end{figure}

The first task is to smooth the noisy power spectra generated from the
N-body simulations. We use the process convolution algorithm described
in~\cite{coyote3}. A process convolution is a mechanism for producing
realizations of a smooth function as a weighted average of a simple
stochastic process. Figure~3 in \cite{coyote3} shows a simple example
with Gaussian variates (the stochastic process) averaged with a
Gaussian smoothing kernel (the weighting scheme).

\begin{figure}[t]
\includegraphics[width=3.5in]{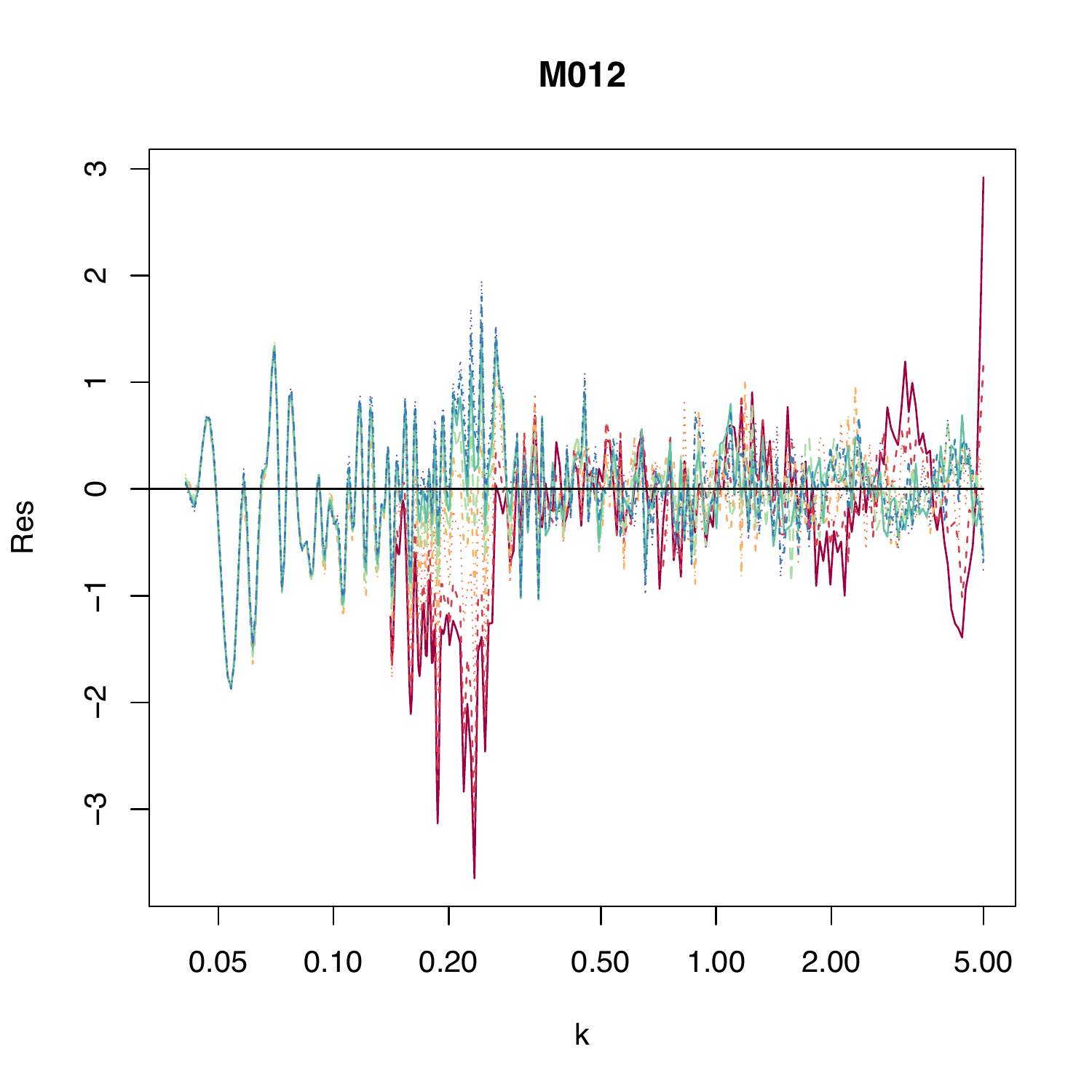}
\caption{Standardized residuals from the high resolution spectra as a
  function of $k$ for M012. The  lines go from yellow to green to blue
  with increasing step (decreasing $z$). These show no obvious trend,
  suggesting that the smooth estimate is a good fit.}
\label{fig:M012resrun}
\end{figure}

As in our previous work, we assume that the unobservable smooth power
spectrum is the result of a two-layer process convolution. The top
layer describes the transformed power spectrum as a process
convolution where Brownian motion realized on a grid is smoothed with
a Gaussian kernel whose kernel width changes over the domain. The
kernel width is described by a the second layer process convolution
which is simply Gaussian variates on a grid smoothed with a Gaussian
kernel. The spectrum computed from each N-body simulation is modeled
as a multivariate Gaussian variable with mean given by the two-layer
process convolution and known diagonal covariance. The unknown smooth
spectrum and a number of nuisance parameters are estimated using a
Markov chain Monte Carlo (MCMC) based approach. Details of the
procedure are provided in~\cite{coyote3}.

The current results are obtained making a slightly different assumption compared to our previous work.  Earlier we assumed, and it appeared to be the case,
that different resolutions had about the same variance for any given
value of $k$ for which a given resolution was unbiased. It now appears
that our current high resolution spectra have smaller variance about
the true spectrum than the low resolution runs. For now, we have used
the larger variance associated with the lower resolution runs in every
case -- as shown below this does not adversely affect the results. In
future iterations, however, we will take this change in variance into
account.

As one example, Figure \ref{fig:M012raw} shows the data from the M012
runs that are used in the estimation procedure for the smooth power
spectra. The dashed gray lines show the TimeRG perturbation theory
results used up to $k = 0.04$Mpc$^{-1}$ for $z < 1$ and up to $k =
0.14$Mpc$^{-1}$ for $z > 1$. The gray dotted lines show the lower
resolution runs that go from the TimeRG perturbation theory results
up to $k = 0.25$Mpc$^{-1}$. The solid black line shows the high
resolution run used from the TimeRG perturbation theory results up to
the maximum value of $k=5$Mpc$^{-1}$. Figure \ref{fig:M012smooth}
shows the estimated smooth spectra for this simulation. M012 is
representative of the results for all parameter settings (the
Appendix~\ref{appendixa} provides a complete list of the sampling
design space).

\begin{table*}
\begin{center}
\caption{Additional Models for Testing\label{tab2}}
\begin{tabular}{ccccccccc}
Model & $\omega_m$ & $\omega_b$ & $\sigma_8$ &   $h$ & $n_s$    & $w_0$ & $w_a$  &$\omega_\nu$ \\
\hline\hline
M038 & 0.1467   &       0.0227  &       0.7325  &       0.5902  &       0.9562  &       -0.8019 &       0.3628  &       0.007077\\
M039 & 0.1209   &       0.0223  &       0.8311  &       0.7327  &       0.9914  &       -0.7731 &       0.4896  &       0.001973\\
M040 & 0.1466   &       0.0229  &       0.8044  &       0.8015  &       0.9376  &       -0.9561 &       -0.0359 &       0.000893\\
M041 &  0.1274  &       0.0218  &       0.7386  &       0.6752  &       0.9707  &       -1.2903 &       1.0416  &       0.003045\\
M042 &  0.1244  &       0.0230  &       0.7731  &       0.6159  &       0.8588  &       -0.9043 &       0.8095  &       0.009194\\
M043 &  0.1508  &       0.0233  &       0.7130  &       0.8259  &       0.9676  &       -1.0551 &       0.3926  &       0.009998\\
M044 &  0.1389  &       0.0224  &       0.8758  &       0.6801  &       0.9976  &       -0.8861 &       -0.1804 &       0.008018\\
\end{tabular}
\end{center}
\end{table*}

\begin{figure}[h]
\includegraphics[width=3.5in]{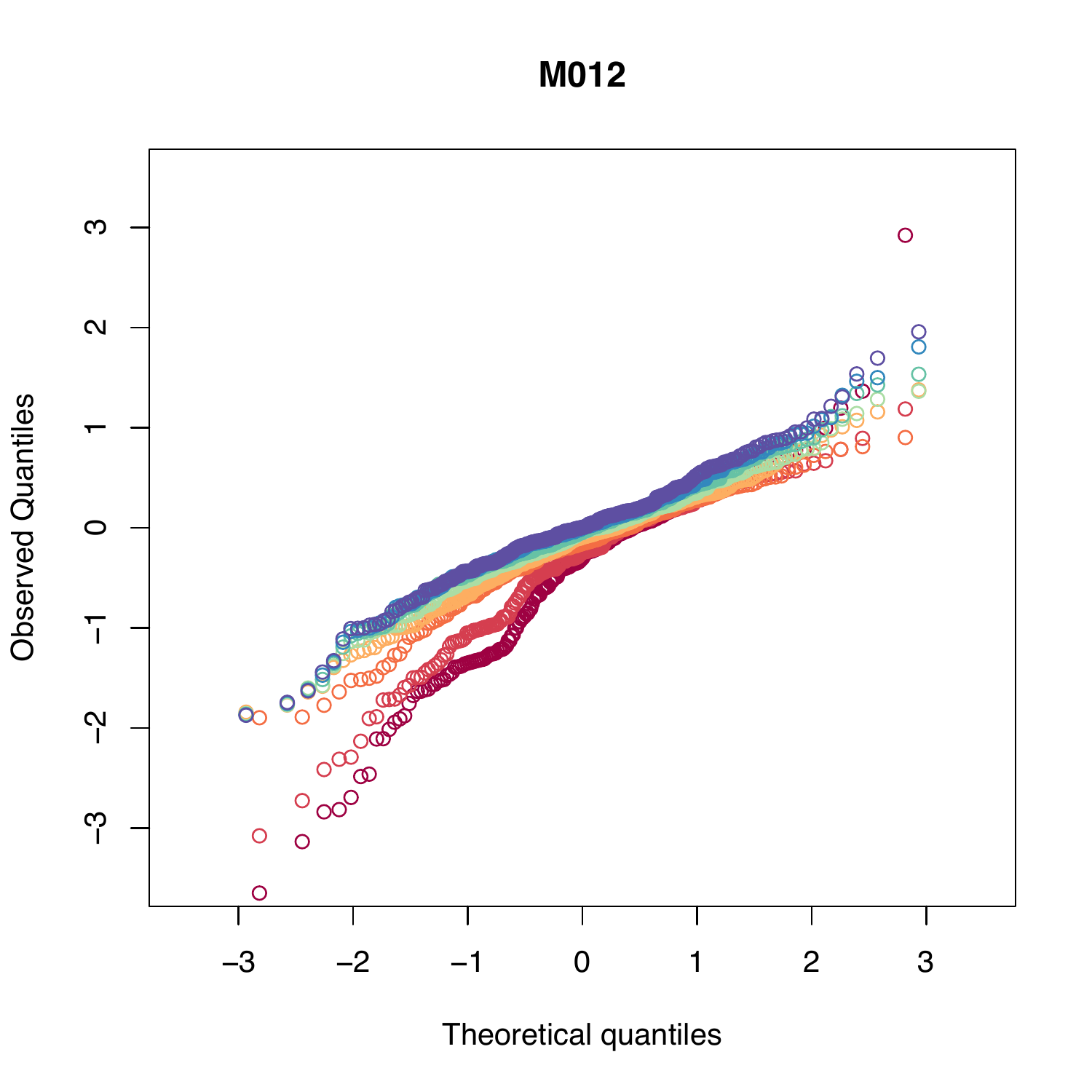}
\caption{Quantile-quantile plot for the standardized residuals from
  the high resolution spectra for  M012. The points go from yellow to
  green to blue with decreasing redshift $z$. Empirical quantiles are
  plotted against the theoretical quantiles of the standard normal
  distribution. The approximately straight lines indicate that the the
  sample is close to normal. The slope suggests that the variance is
  somewhat smaller than expected, as discussed in the text.} 
\label{fig:M012qq}
\end{figure}

Figures~\ref{fig:M012resrun} and \ref{fig:M012qq} show some
diagnostics for the process convolution fit. Both of these plots
consider the standardized residuals for the high resolution run for
cosmology M012 (the other cosmologies and resolutions lead to similar
conclusions). The standardized residuals are computed in the following
manner. First, the smoothed process convolution estimate is subtracted
off across $k$. If the process convolution process has provided a good
estimate for the mean, the residuals should now have zero mean across
$k$, that is the mean taken across $k$ should be close to zero. Next,
each residual is divided by the standard deviation, the square root of
the variance, of the raw data at each $k$. This variance changes over
$k$ in a log-linear fashion, i.e., the log of the variance decreases
linearly with the logarithm of $k$. This behavior is used as part of
the process convolution procedure (see \citealt{coyote3} for details.)
If this variance prediction is correct, the resulting standardized
residual should have variance one. If there is little correlation, the
collection of standardized residuals should resemble an independent
sample from the standard normal distribution.

Figure~\ref{fig:M012resrun} presents evidence that the process
convolution works well in capturing the mean structure. The
residuals are centered on zero across $k$ and there are no major
trends in the data, supporting the unbiased nature of the mean
estimate. (There is some evidence of oscillatory behavior at high $k$,
which might indicate that our process convolution mean is not flexible
enough or might arise for some other cause; either way, it is not
quantitatively significant.) Here we see some confirmation of the
aforementioned fact that the high resolution runs have smaller
variance than the low resolution run. Most of these residuals are
between -1 and 1, which is too small to match our assumption that
these residuals should resemble draws from a standard normal (which
would produce numbers mostly between -3 and 3). Figure
\ref{fig:M012qq} tests our distribution of Gaussianity. Here, the
empirical quantiles of the standardized residuals (basically the
sorted residuals) are plotted against the theoretical quantiles of the
standard normal distribution. The colors match Figure
\ref{fig:M012resrun}. The straight lines indicate that the residuals
do appear to be Gaussian and are relatively uncorrelated. The slope of the
lines is related to the variance. The fact that these are less than
unity is another indication that the variance of these residuals is
smaller than expected. This conclusion about the variance is not
detrimental as every indication is that we are estimating the smooth
spectra well and the errors are fairly Gaussian.

\subsection{Emulation}

\begin{figure}
\includegraphics[width=3.5in]{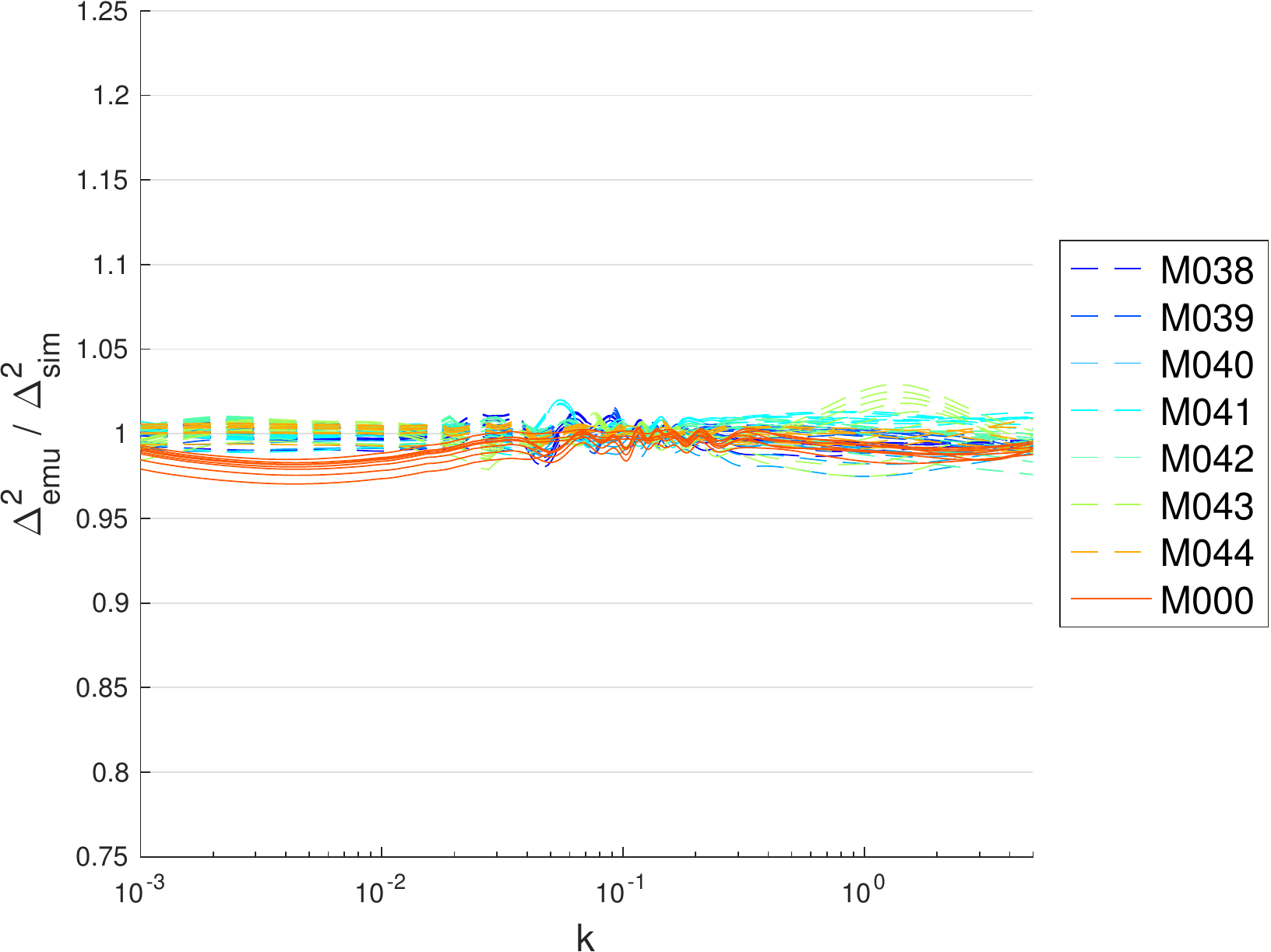}
\caption{Total matter power spectrum predictions for the
  cosmologies M038-M044 (completed runs from the next design stage,
  not used in the current work) and the best fit WMAP7 cosmology M000. The
  former are a hybrid of out-of-sample and cross-validation predictions
  as the TimeRG perturbation theory-only runs for these cosmologies
  were held-out on each prediction. The model, M000, is a true test set.
  The comparison implies that the error is less than 3\%.} 
\label{fig:ptot_test}
\end{figure}
\begin{figure}[t]
\includegraphics[width=3.5in]{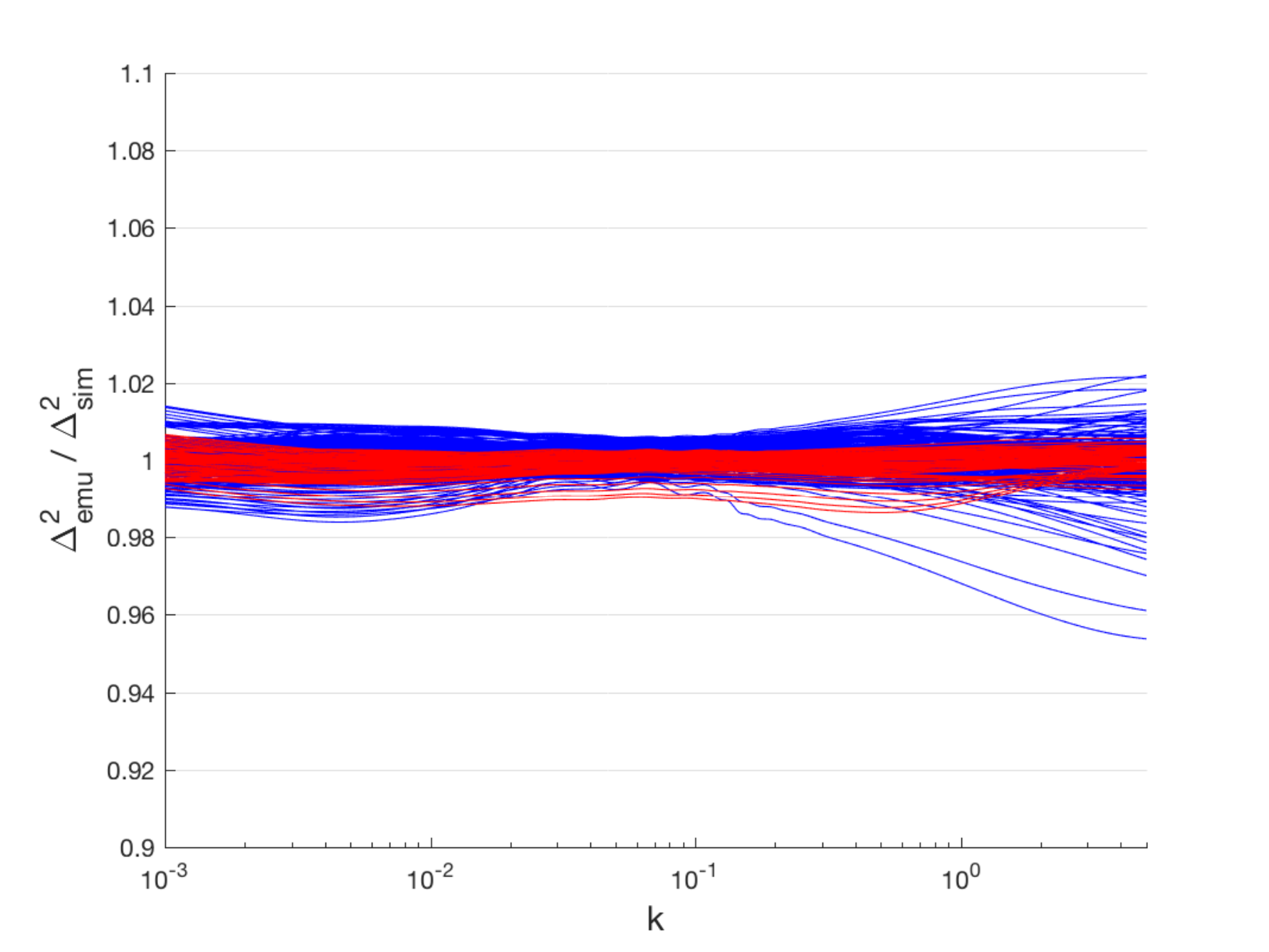}
\caption{Total matter power spectrum predictions for the training set.
  Red lines show simulations with $\omega_\nu=0$ (M000-M010) while
  blue lines show simulations with non-zero neutrino masses
  (M011-M036). Normally, an emulator is expected to interpolate these
  exactly, but the inclusion of weights from the partial runs makes
  this difficult. The resulting emulator behaves somewhat more like
  standard regression which does not interpolate, but minimizes error.
  These results suggest that the error may be as high as 5\% in some
  cases. As the number of runs increases, this issue will be resolved (see text).} 
\label{fig:ptot_leavein}
\end{figure}

The follow-on task is to construct the emulator from smooth estimates
of the matter power spectra. As in the extension of the original
Cosmic Emulator in \cite{emu_ext}, we also have a number of partial
power spectra using the TimeRG perturbation theory approach. In fact,
we have results from TimeRG perturbation theory for the complete
design (111 models). To build the emulator, we follow a version of the
basic plan from \cite{emu_ext}. Our goal is to predict the
multivariate power spectrum from an N-body simulation as a function of
the eight input parameters. As detailed in \cite{coyote3}, the first
step is to standardize the simulation outputs by centering and
scaling, and then projecting them on to an empirical basis computed
via SVD (i.e., principal components or empirical orthogonal
functions). This process discovers the directions of greatest
variation in the high-dimensional outputs and reduces the modeling to
these dimensions. The basis weights are then modeled as functions of
the simulation inputs using Gaussian processes.

In this case, we have TimeRG perturbation theory results for the
entire 111 run design and complete, smoothed spectra from N-body
results over the first 36 runs in the design. The 36 complete runs are
used to compute the mean vector, the scaling factor, and 35 basis
vectors. The complete spectra are centered, scaled, and projected onto
the basis to obtain their weights. The partial TimeRG perturbation
theory power spectra (up to the $k$ values described in the
description of the smoothing) are centered using the relevant portion
of the mean vectors, scaled, and then projected on to the relevant
part of the basis vectors to obtain their weights. However, the
partial power spectra are only projected onto the first 7 basis
vectors. This number was chosen by comparing the weights from the
partial and complete runs. Beyond 7 basis vectors, the weights from
the partial runs begin to differ visually from the weights from the
complete runs when plotted against the eight input parameters. As a
result, using the partial run weights will actually begin to degrade
the performance of the emulator. All of the weights, from both
complete and partial spectra are put together to estimate the Gaussian
process emulator. See \cite{coyote3, emu_ext} for details on the
estimation via an MCMC based approach.

Figure~\ref{fig:ptot_test} shows test results of the emulator fit for
the total matter power spectrum emulator, $P_{tot} = (P_{cb}^2 +
P_{\nu}^2)^{1/2}$. The dashed lines show the results for spectra from
design points M038-M044, which are completed runs from the next stage
of the emulator lattice design. The tests are done by holding out the
partial TimeRG perturbation theory results for these runs and
predicting the complete spectra. The maximum error is about 3\% and
most are below 2\%. The solid lines are the results for the best fit
cosmology M000. This is a true out of sample test. At low $k$ the
error reaches its maximum, with one redshift showing about a 2.5\%
error. Figure~\ref{fig:ptot_leavein} shows the results from predicting
the training set. Typically, emulators are expected to interpolate the
training set, but that is not true in the current case. It seems
likely that the emulator has difficulty interpolating the weights from
the TimeRG perturbation theory-only results. The resulting fit becomes
more like standard regression where the data is not interpolated, but
errors are minimized. This issue shows up most prominently at high $k$
where the TimeRG perturbation theory-only runs provide no direct
information. Still, the worst-case error is only 5\% with the vast
majority of errors under 2\%. (These numbers are consistent with the
linear theory tests carried out in \citealt{heitmann15}.) As the
number of complete runs continues to increase in later releases, we
anticipate that this issue will disappear; tests using the linear
theory results from \cite{heitmann15} are consistent with this
expectation. Overall, the emulator performs very well despite only 36
complete sets of spectra in eight dimensions. We also measured the
errors in just the the baryon-dark-matter component, $P_{cb}(k)$,
finding very similar results.

\section{Comparison with other approaches}
\label{comparison}

In this section we provide some comparisons with alternative
approximate prediction methods. Since most other groups have addressed
either neutrinos or a dynamical dark energy equation of state but not
both (as done here), we divide our tests accordingly. In each of the
following subsections we compare the alternative approaches to our
full simulations, if an appropriate model is available. In addition,
we use a set of new models that are not in the simulation design to
compare the emulator with the other prediction schemes if those
schemes do allow variation of all eight parameters. We carry out our
comparisons for the two extreme redshifts, $z=0$ and
$z=2.02$.

\begin{figure}[t]
\includegraphics[width=3.5in]{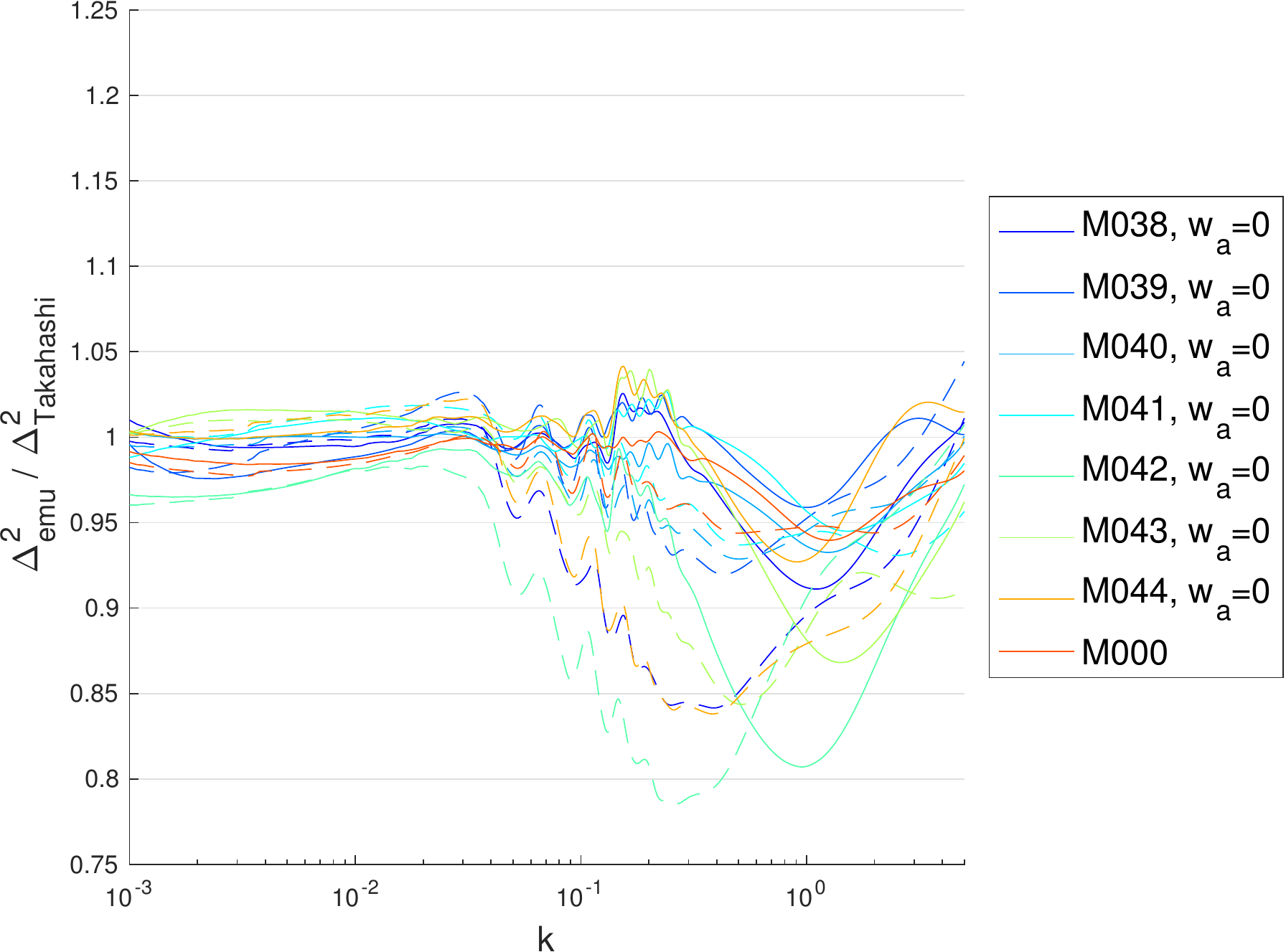}
\caption{Ratio of predictions from the new emulator with those
  obtained using the method of \cite{takahashi12}. Solid lines show
  results at $z=2.02$ while dashed lines show results at $z=0$.  The
  methods match very well at low k (at the 2-3\% level), but differ by
  20\% at high $k$ for the higher neutrino masses with $\omega_\nu >
  0.007$ (M038, M042, M043, M044). For the models with $\omega_\nu <
  0.031$, the agreement is better than 7\% over most of the $k$-range
  investigated.} 
\label{fig:takacomp}
\end{figure}

\subsection{Neutrino Predictions}

For the case of power spectrum predictions including neutrinos, we
study the Halofit approach by \cite{takahashi12}, which was augmented
with a neutrino term by~\cite{bird}. \cite{takahashi12} improved the
Halofit model originally developed by~\cite{smith03} around the
$\Lambda$CDM model by adding a set of sixteen high-quality
gravity-only simulations. Six of those models were chosen around the
best-fit WMAP results from different years. The other ten were at the
same design points as the first ten models from the original Coyote
Emulator~\citep{coyote3} (the simulation results agreed with those run
for the Coyote Emulator mostly within $3\%$). These additional simulations
allowed them to include a constant equation of state parameter $w$ as
a new cosmological parameter. Next, they refitted their parametric
model (adding additional parameters) to achieve an accuracy at the
5-10\% level out to $k\le 10h$Mpc$^{-1}$. Based on this work,
\cite{bird} added a neutrino component to Halofit with a new set of
simulations covering neutrino  masses between $0.15\le \sum m_{\nu}
\le 0.6$eV. Their neutrino treatment is particle-based, with the
neutrinos modeled as a separate species, albeit at lower force
resolution than the dark matter particles. In order to avoid problems
due to large neutrino velocities, they started the simulations as late
as $z_{in}=24$ for the lightest neutrinos, using the Zel'dovich
approximation. This leads to systematic inaccuracies in the power
spectrum at the few percent level as shown in~\cite{coyote1} and
\cite{schneider}. (In both papers effects at the 2-3\% level were
shown with a starting redshift $z_{\rm in}=50$ at $k\sim
1h$Mpc$^{-1}$, extrapolating these results would suggest a 5\% error
due to the late start alone, and even more at higher $k$). In addition,
the small volumes and limited mass resolution (512$^3$ particles)
further degrade the accuracy of the simulations. All these effects
combined will lead to systematic errors and scatter in the power
spectrum, in particular on small length scales. The neutrino-augmented
results are available in the latest CAMB release and have been updated
over time~(\citealt{camb}).

Figure~\ref{fig:takacomp} shows a comparison of our new emulator with
the \cite{takahashi12} implementation. The $\Lambda$CDM model (M000,
brown line) agrees with our emulator at the 5\% level at $z=0$ which
is in agreement with our previous findings in \cite{emu_ext} for the
same model (see Fig.~11 in that paper. Note that in the extended
emulator paper the ratio is taken with respect to the simulation,
meaning that the $y$-axis in that paper is the inverse from what we
show in Figure~\ref{fig:takacomp} here). Our finding of very similar
agreement with~\cite{takahashi12} with the new emulator for M000
stresses the excellent agreement between {\sc Gadget-2} and HACC (the
original Coyote Emulator papers were based on {\sc Gadget-2}
simulations, while the new simulations have been carried out with
HACC, agreements are well within the sub-percent level). Similar results
comparing {\sc Gadget-2} and HACC were also reported in the HACC code
paper by~\cite{habib16}.

For all models, the agreement on large scales (small $k$,
$k<0.02$Mpc$^{-1}$) is at the 1-2\% level at both redshifts, $z=0$ and
$z=2.02$, demonstrating that our use of a $k$-dependent correction
factor for the growth function works very well. In the quasi-linear to
nonlinear regime the agreement between the Halofit approach and the
new emulator varies between 5\% up to 20\%. This is again consistent
with our previous findings in \cite{emu_ext}, Figure~12, where for
some cosmological models, the differences for the power spectrum
prediction between Halofit and the extended emulator were as large as
15\% over a similar $k$-range ($0.1{\rm Mpc}^{-1} < k < 1{\rm
  Mpc}^{-1} $). Contemplating this level of error in Halofit is
discomforting in the context of using the matter power spectrum to
obtain cosmological constraints, since these deviations of around $\sim
10-15\%$ occur in the range of scales typically accessed by
measurements of the cosmic shear power spectrum.

Finally, we emphasize that as shown in our previous
work by~\cite{upadhye14}, the agreement of our simulations including
neutrinos with a TimeRG-based perturbative approach was better than
2\% at $\sim 0.2$Mpc$^{-1}$ at $z=2$ and $\sim 0.1$Mpc$^{-1}$ at
$z=0$, for values of $\omega_\nu$ as high as $0.01$. Given these
results, in combination with the findings in~\cite{castorina15}
discussed in Section~\ref{sec:neutrinos}, the differences in Halofit
and the new emulator are apparently due to the general inaccuracy of
Halofit away from $\Lambda$CDM models rather than to the different
neutrino treatment applied.

\subsection{Dynamical Dark Energy Equation of State Predictions}

\begin{figure}[b]
\includegraphics[width=3.5in]{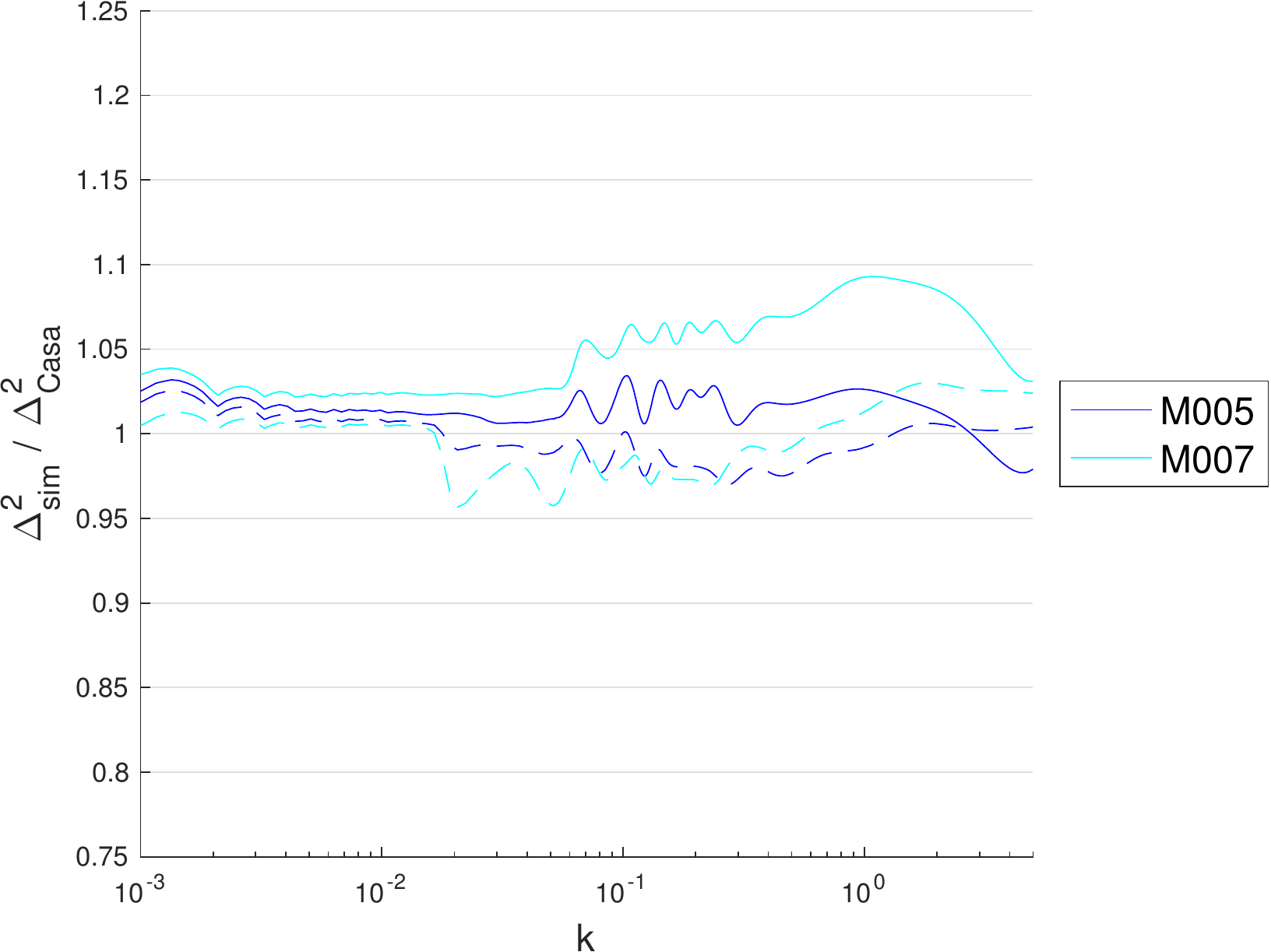}
\caption{Ratio of simulation results with those obtained using the
  method of \cite{casarini} for M005 and M007. We compare at $z=2$
  (solid line) and $z=0$ (dashed line). For M007 we have $w_a=-1.0$
  and for M008 we have $w_a=0.4333$. The values for $\omega_m$ and
  $\sigma_8$ are higher for M007, leading to stronger nonlinear effects.
  The \cite{casarini} approach leads to $<5\%$ inaccuracy for M005
  and $<10\%$ for M007.} 
\label{fig:casacompsim}
\end{figure}

\begin{figure}[t]
\includegraphics[width=3.5in]{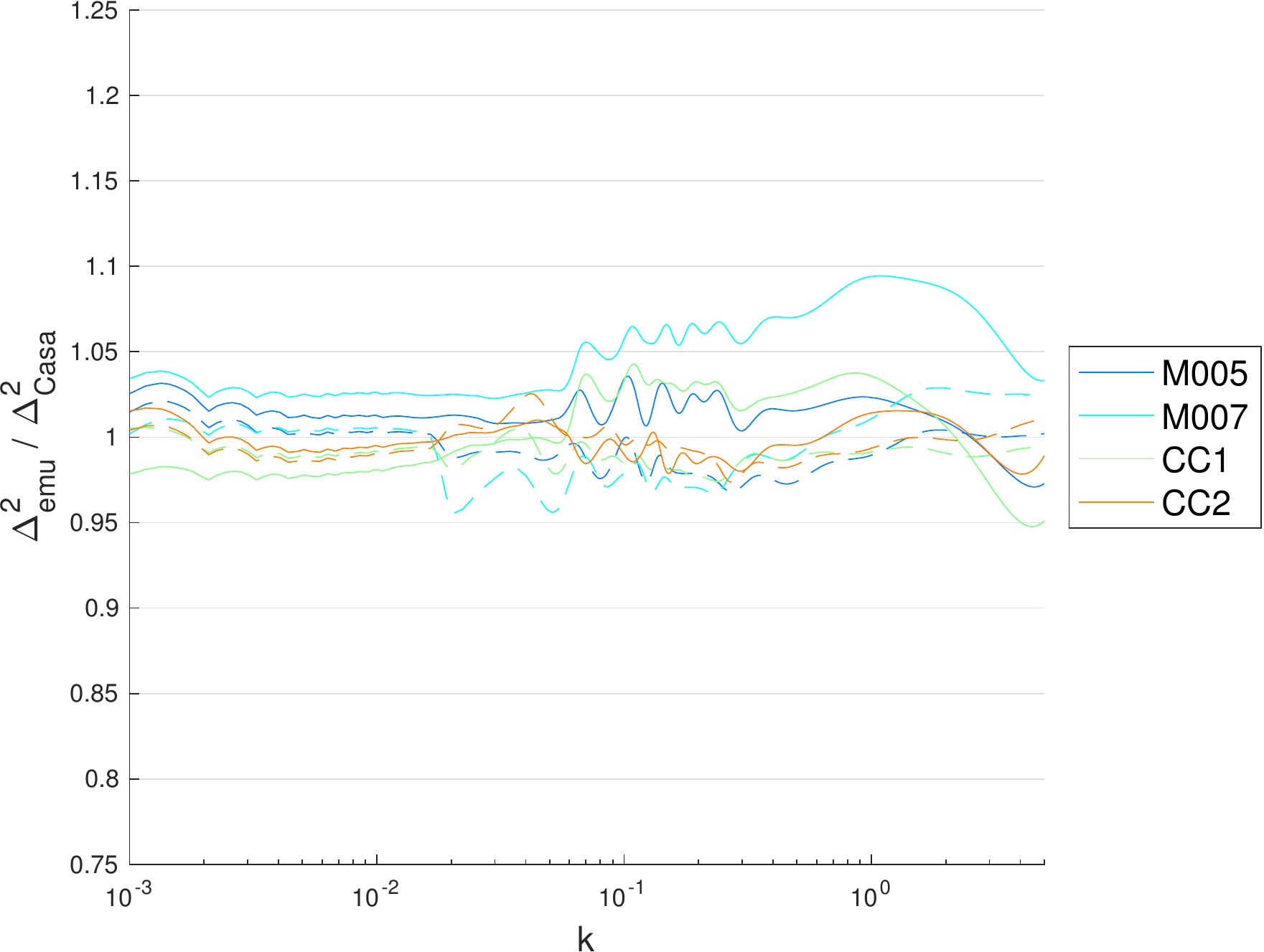}
\caption{Ratio of predictions from the new emulator with those
  obtained using the method of \cite{casarini}. CC1 and CC2 are
  cosmologies selected in addition to M005 and M007 for this
  comparison. The inputs are $\omega_m = 0.14205$, $\omega_b =
  0.02225$, $\sigma_8 = 0.83$, $h = 0.6727$, $n_s = 0.9645$ and
  $\omega_{\nu} = 0$ for both CC runs. For CC1, we choose $w_0 = -1.0$
  and $w_a = 0.5$ and for CC2, $w_0 = -1.0$ and $w_a = -0.5$. For all
  cosmologies, we compare at $z=2$ (solid line) and $z=0$ (dashed
  line). The methods match reasonably well, agreeing within 10\%
  errors. The disagreement between the emulator and the
  \cite{casarini} result for M005 and M007 is at the same level as the
  disagreement of direct comparison with the simulations in
  Figure~\ref{fig:casacompsim}, consistent with our emulator test
  shown in Figure~\ref{fig:ptot_leavein}.}
\label{fig:casacomp}
\end{figure}

Next we compare our results with the work of~\cite{casarini}. 
Based on our
earlier emulator work in~\cite{emu_ext}, these authors developed a
prediction for $(w_0,w_a)$ cosmologies by introducing an effective
constant equation of state that captures the influence of a
time-varying dark energy equation of state to sub-percent accuracy.
This idea was introduced in~\cite{francis07} and is based on the
assumption that cosmologies beyond $\Lambda$CDM can be mapped back to
$w$CDM models by requiring both models to have the same distance to
last scattering and the same values of H$_0$ and energy densities,
$\Omega_{{m,r,b},0}$ at $z =$ 0.  This has the effect of tuning the
growth in constant $w_0$ models to match the $(w_0,w_a)$ models of
interest for some new value of $\sigma_8$. The new value of $\sigma_8$
is constrained by the chosen values of $(w_0, w_a)$ and within the
context of the Coyote emulator, this limits the space of allowable models
because of the parameter range of the design.

Based on the above general idea and the results from~\cite{emu_ext}
for $w$CDM models,~\cite{casarini} deliver predictions for the
nonlinear power spectrum for dynamical dark energy models for scales
of $1< k < 1.5$ and between redshift $0\le z\le 3$ at high accuracy.

Before we show a comparison of the \cite{casarini} approach with the
emulator, we compare two of our smoothed power spectra from the
simulations for M005 and M007 directly with their prediction in
Figure~\ref{fig:casacompsim}. By testing against the smoothed input
power spectra as well, we are able to distinguish between various
sources of error in Figure~\ref{fig:casacompsim}, that is, whether the
discrepancy, should there be any, is due to the assumptions of the
Casarini et al. model or the predictive power of the Gaussian Process
modeling. The solid line shows results at $z=2$ while the dashed
lines show results at $z=0$. The results for $z=0$ are in agreement at
a level better than 5\%, for M005 at both redshifts. For M007 at $z=0$ the agreement is also excellent (below 5\%), for $z=2$ it degrades slightly but stays well under 10\%. Next,
Figure~\ref{fig:casacomp} shows a comparison of the emulator with
\cite{casarini} for the same models and two additional models, CC1 and
CC2 that were not part of our original design (the cosmological
parameters for CC1 and CC2 are listed in the figure caption). We have
chosen to include these entirely new models in our analysis because
the behavior of $\sigma_8(z)$ cannot be chosen independently of the
other cosmological parameters. This meant that we could only generate
predictions for two models in our testing set, without massive
neutrinos, staying within the Casarini et al. framework

Again, the comparisons are carried out at $z=0$ and $z=2$. At
large scales ($k<0.02$Mpc$^{-1}$) the agreement is excellent, at the
1-2\% level. In the quasi-linear to nonlinear regime, the agreement of
the emulator and \cite{casarini} is very similar to the agreement with
respect to the simulations themselves, better than 5\% for all models
at $z=0$ and well below 10\% for all models at $z=22$. This level
of agreement shows that this approach works quite well.

\subsection{Eight-Parameter Model Predictions} 

In their work, \cite{mead16} provide new power spectrum predictions,
covering not only neutrinos but also dynamical dark energy and
modified gravity models. Their approach is based on re-deriving the
different contributions to the halo model. The new prediction scheme
is supposed to be valid for $k < 10h$Mpc$^{-1}$ at the few percent
level accuracy for most models. They arrive at this conclusion by
comparing their results to a range of simulations. This is the only
approach for which we can compare our full emulator and our simulation
results directly. We use the additional seven models for this
comparison that were also employed to investigate the accuracy of our
emulator, as given in Table~\ref{tab2}.

\begin{figure}[t]
\includegraphics[width=3.5in]{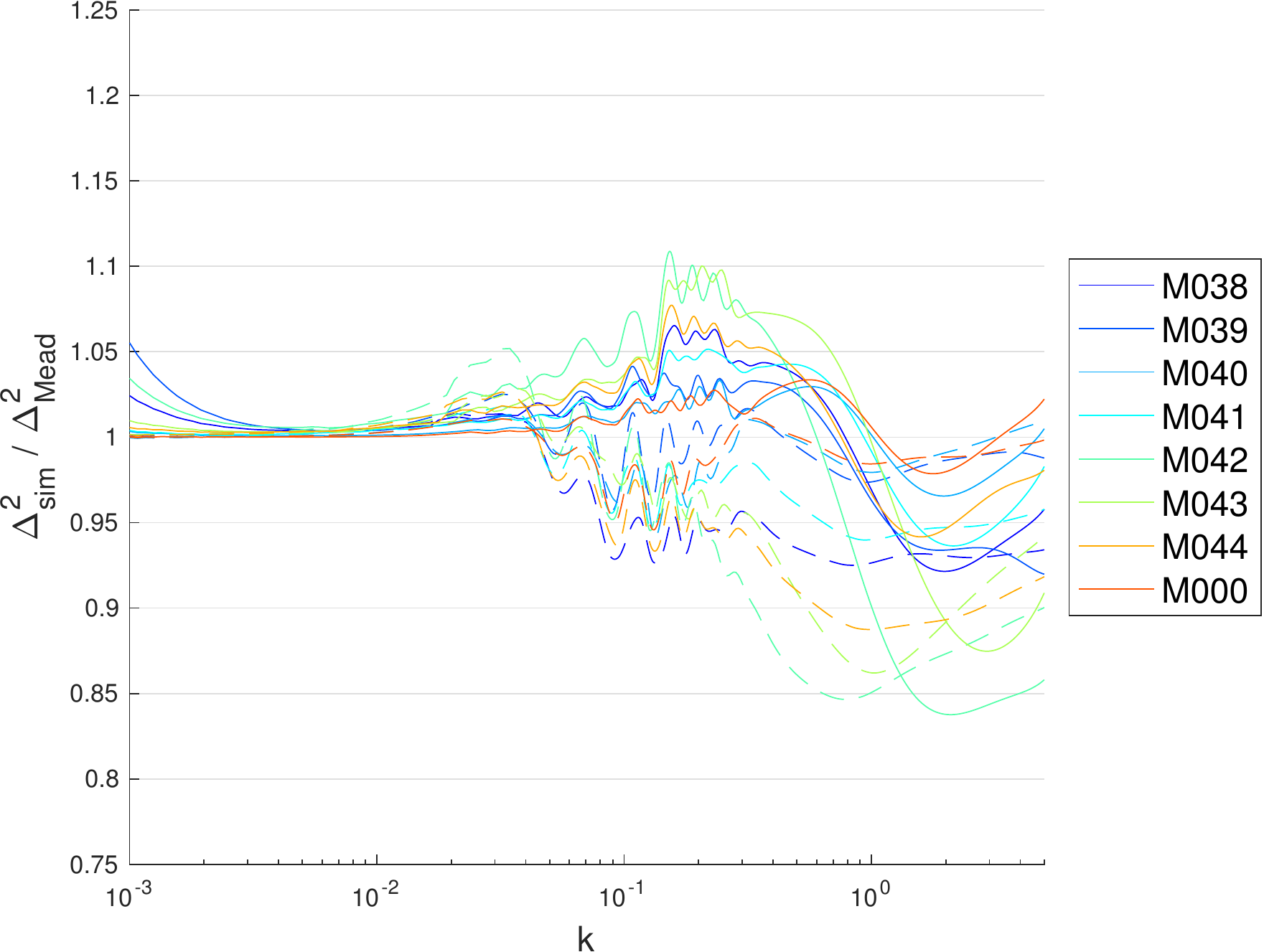}
\caption{Ratio of predictions from our smoothed simulation results
  with those obtained using the method of \cite{mead16} for M038-M044
  given in Table~\ref{tab2} and M000, our $\Lambda$CDM cosmology.
As in previous plots, dashed lines show the results at $z=0$ and solid
lines at $z=2.02$. The agreement for M000 is basically perfect on
large scales and at the 2\% level in the nonlinear regime. For
M038-M044 the agreement is good over most of the $k$-range (at the
5-10\% level) but discrepancies can be as large as 15\%.} 
\label{fig:meadsim}
\end{figure}

\begin{figure}[t]
\includegraphics[width=3.5in]{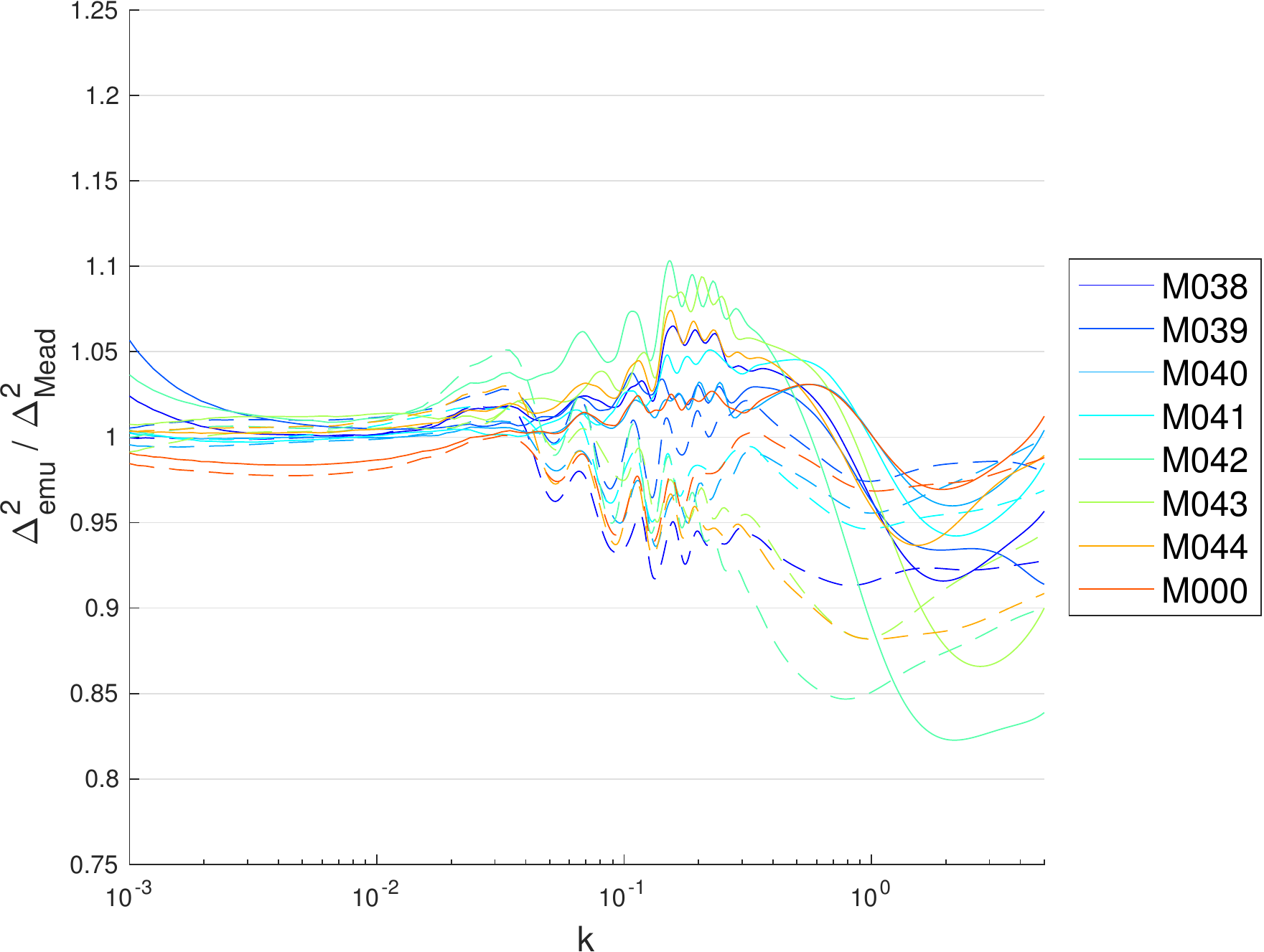}
\caption{Ratio of predictions from the new emulator with those
  obtained using the method of \cite{mead16} for the same models shown
  in Fig.~\ref{fig:meadsim}. The methods match roughly at the 3\%
  level at low $k$ for most models,  but differ by more than 15\% at
  high $k$ consistent with the comparison to the simulations
  themselves shown in Figure~\ref{fig:meadsim}.}
\label{fig:meadcomp}
\end{figure}

We show the comparison of the \cite{mead16} fit with our smoothed
simulations directly in Figure~\ref{fig:meadsim} for M000 and
M038-M044, given in Table~\ref{tab2}. The agreement for the
$\Lambda$CDM cosmology is excellent, basically perfect on large scales
up to $k\sim 0.04$Mpc$^{-1}$ and at the $2-3\%$ in the nonlinear
regime. The models with a varying dark energy equation of state show some
disagreement on the very largest scales, which is most likely due to
our different implementation of $(w_0,w_a)$ cosmologies in CAMB than
in the version that was used by \cite{mead16}. For relevant details
the reader is referred to our previous work (\citealt{upadhye14})
where we provide a description of our implementation and a publicly
available CAMB version. In the quasi-linear regime, the agreement is
around 5-10\% while at $k\sim 1$Mpc$^{-1}$ some models show
differences of up to 15\%.  

Figure~\ref{fig:meadcomp} shows the ratio of the new emulator with
respect to the \cite{mead16} fit for the same models. The results are
very similar to the comparison to the smoothed simulations as to be
expected from Fig.~\ref{fig:ptot_test} (obviously, taking the ratio of
Figures~\ref{fig:meadsim} and \ref{fig:meadcomp} would lead back to
the results of Figure~\ref{fig:ptot_test}). On large scales, we see
slightly poorer agreement compared to the comparison to the
smoothed simulations, but still at the $2\%$ level. On small scales,
the agreement is very similar to the direct comparison with the
simulations -- for some models around $5\%$ while for others closer to
$15\%$.

\subsection{Comparison with the Extended Emulator}

\begin{table*}
\begin{center}
\caption{Extended Cosmic Emu Comparison Cosmologies\label{tabFE}}
\begin{tabular}{ccccccccc}
Model & $\omega_m$ & $\omega_b$ & $\sigma_8$ &   $h$ & $n_s$    &
$w_0$ & $w_a$  &$\omega_\nu$ \\ 
\hline\hline
FC1 & 0.15110 & 0.02217 & 0.81110 & 0.8167 & 1.0280 & -1.09038 & 0 & 0 \\
FC2 & 0.15110 & 0.02217 & 0.82813 & 0.8167 & 1.0280 & -1.19484 & 0 & 0 \\
FC3 & 0.12000 & 0.02306 & 0.70000 & 0.6833 & 1.0060 & -0.74838 & 0 & 0 \\
FC4 & 0.12000 & 0.02306& 0.68981 & 0.6833 & 1.0060 & -0.71040 & 0 & 0 \\
FC5 & 0.14205 & 0.02225 & 0.83000 &0 .6727 & 0.9645 & -0.84095 & 0 & 0 \\
FC6 & 0.14205& 0.02225 & 0.81830 &0 .6727 & 0.9645 & -0.78951 & 0 & 0 \\
FC7 & 0.14205 & 0.02225 & 0.83000 &0 .6727 & 0.9645 & -1.12210 & 0 & 0 \\
FC8 & 0.14205& 0.02225 & 0.83757 &0 .6727 & 0.9645 & -1.18190 & 0 & 0 \\
\end{tabular}
\end{center}
\end{table*}

\begin{figure}[h]
\includegraphics[width=3.5in]{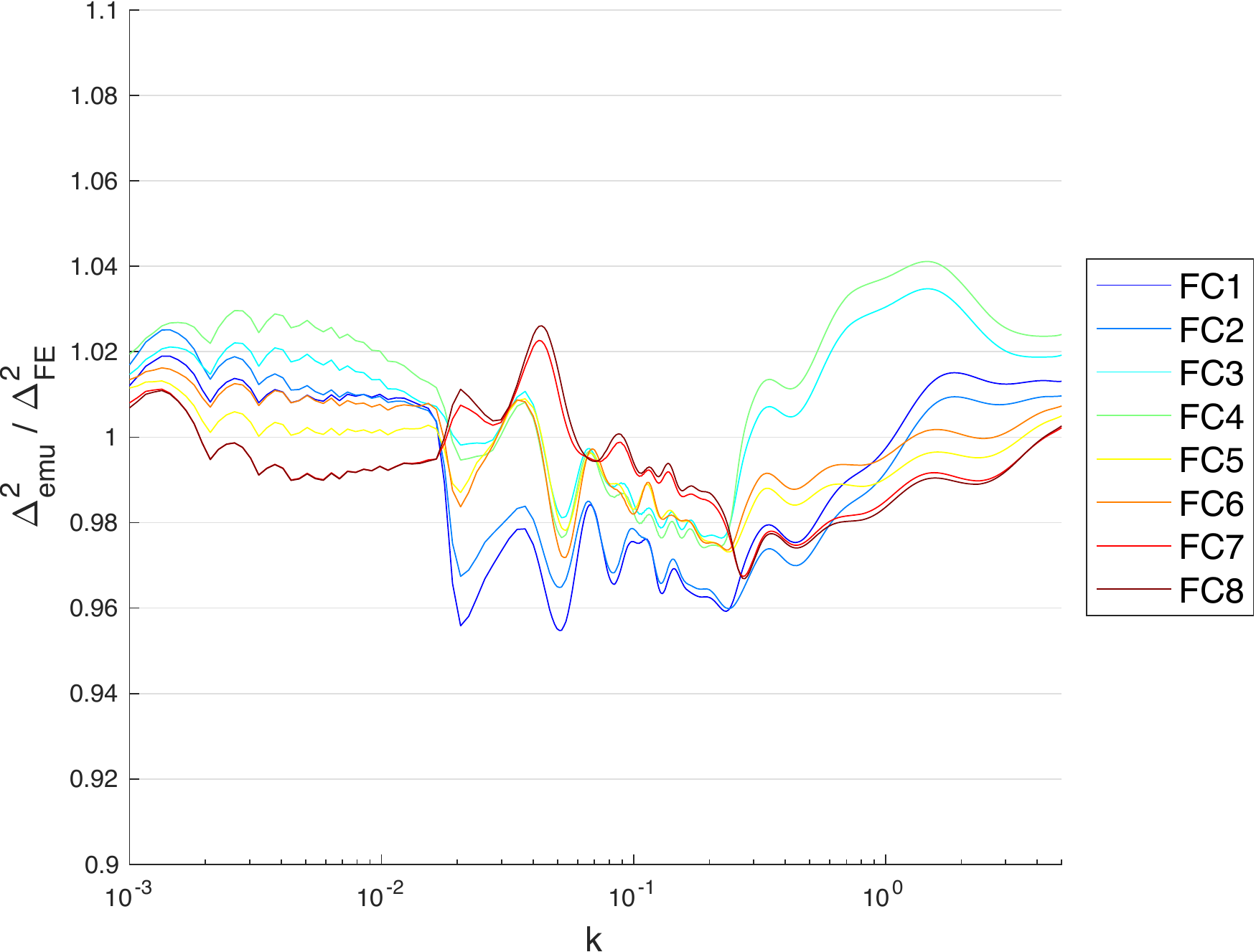}
\caption{Ratio of predictions from the new emulator with those
  obtained using the extended Cosmic Emulator \citep{emu_ext}. The
  cosmological parameters for this comparison are given in Table
  \ref{tabFE}. For all cosmologies, we compare at $z=0$. The methods
  match within their separately estimated errors, with disagreements at
  most at $4\%$.}   
\label{fig:frankencomp}
\end{figure}

Finally, as a last check, we compare our earlier, `Coyote Extended'
emulator, developed in~\cite{emu_ext} with our new emulator, keeping
$w_a=\omega_\nu=0$. We list the models used for this test in
Table~\ref{tabFE}. Several of the values used are close to the edge of
our parameter design, making this test rather stringent.
Figure~\ref{fig:frankencomp} shows the ratio of the new emulator over
the extended emulator from our previous work. We show results for
eight models at redshift $z=0$. The agreement for most models is at
2\% with two instances showing disagreement up to 4\%. Given the
estimated accuracy of our current emulator at approximately 4\% and of
the previous emulator at the $\sim 5\%$ level this agreement is within
the expected limits.

\section{Summary and Outlook}
\label{conclusion}
We introduce a new cosmic emulator for the matter power spectrum that
covers eight cosmological parameters, and spans a redshift range from $0\le
z\le 2$ and wave numbers out to $k\sim 5$Mpc$^{-1}$. We achieve an accuracy at the 4\%
level (better for most models) over the full $k$-range and all eight
parameters while using a sampling space of just 36 cosmological
models. 

The parameter sampling approach for designing the simulation suite is described in in~\cite{heitmann15}. Because this scheme has demonstrated convergence properties, the accuracy of the emulator can be systematically improved by adding
more simulations at well-defined points reaching close to $1\%$ with
about 100 evaluation points in the eight-dimensional space; the next
set of 26 simulations is currently being analyzed.  The internal
accuracy tests presented in this paper are consistent with the
estimates from the linear theory-based test presented in
~\cite{heitmann15} and with the error estimates of a previously
constructed emulator that relied on a different set of
simulations~\citep{emu_ext}. These positive results are an important
consistency check for our planned further improvement of the error bounds.

We have compared our results to other predictions for the nonlinear
power spectrum by a number of authors. The agreement in the linear
regime is excellent as is to be expected. The agreement degrades for
models away from $\Lambda$CDM in the quasi-linear to nonlinear regime
and we find differences at the $5-10\%$ for most models but up to
$15-20\%$ depending on the model and redshift investigated and
prediction scheme used (only \citealt{mead16} provide predictions over
the eight cosmological parameters that we investigate here, all other
approaches only treat a subset of the parameters). The agreement we
find between different methods is consistent with our evaluation in
the extended Coyote emulator, presented in~\cite{emu_ext}, where we
found differences between the emulator predictions and, e.g., Halofit
at the 20\% level for some models.

The simulation suite presented here lends itself to many more
investigations. We are currently building a large set of emulators for
quantities such as the halo mass function, redshift space distortions,
and halo correlation and galaxy correlation functions and bias
functions. As emulation accuracies continue to improve, addition of `post-processing' modules for, e.g., baryonic effects and galaxy modeling, will also become easier to implement in a robust fashion.

In future, we expect emulators and observations to co-evolve. There will likely be a greater emphasis on cross-correlation-based probes; also as measurements squeeze the parameter space priors, the quality of emulation will improve significantly.

\begin{acknowledgments}

We thank Alexander Mead and Tim Eifler for many useful conversations.
KH and SH thank the Aspen Center for Physics, which is supported by
National Science Foundation grant PHY-1066293, where part of this work
was carried out.  

Part of this research was supported by the DOE under contract
W-7405-ENG-36. Argonne National Laboratory's work was supported
under the U.S. Department of Energy contract DE-AC02-06CH11357.
Partial support for HACC development and for this work was provided by the Scientific
Discovery through Advanced Computing (SciDAC) program funded by the
U.S. Department of Energy, Office of Science, jointly by Advanced
Scientific Computing Research and High Energy Physics.
   
This research used resources of the ALCF, which is supported by DOE/SC
under contract DE-AC02-06CH11357 and of the Oak Ridge Leadership
Computing Facility at the Oak Ridge National Laboratory, which is
supported by the Office of Science of the U.S. Department of Energy
under Contract No. DE-AC05-00OR22725.  
\end{acknowledgments}

\appendix
\label{appendixa}

In this Appendix we list all the cosmological models that have been
used in the paper to construct and test the new emulator.

\begin{table*}
\begin{center}
\caption{Design\label{tab1}}
\begin{tabular}{ccccccccc}
Model & $\omega_m$ & $\omega_b$ & $\sigma_8$ &   $h$ & $n_s$    & $w_0$ & $w_a$  &$\omega_\nu$ \\
\hline\hline
M000 & 0.1335 & 0.02258 &  0.8 & 0.71 & 0.963 & -1.0 & 0.0 & 0.0 \\
M001 & 0.1472  &  0.02261& 0.8778 & 0.6167 & 0.9611 &  -0.7000&  0.67220 & 0.0  \\
M002 &  0.1356 &  0.02328 &  0.8556& 0.7500  & 1.0500 &  -1.0330 & 0.91110 &  0.0  \\
M003 &   0.1550&  0.02194 &0.9000 & 0.7167 & 0.8944 &  -1.1000 & -0.28330 &  0.0  \\
M004 &  0.1239 &    0.02283 & 0.7889& 0.5833 & 0.8722 & -1.1670 & 1.15000 & 0.0   \\
M005 &  0.1433 & 0.02350 &  0.7667&0.8500  & 0.9833 & -1.2330 & -0.04445 &  0.0  \\
M006 &   0.1317& 0.02150 & 0.8333 & 0.5500  &0.9167  &  -0.7667& 0.19440 &   0.0 \\
M007 &   0.1511&  0.02217& 0.8111 & 0.8167 & 1.0280 & -0.8333 &  -1.00000 &  0.0  \\
M008 &  0.1200 & 0.02306 & 0.7000 & 0.6833 & 1.0060  & -0.9000 & 0.43330 &  0.0  \\
M009 &   0.1394& 0.02172 & 0.7444 &  0.6500  &  0.8500 & -0.9667 &  -0.76110& 0.0  \\
M010 &  0.1278 & 0.02239 &  0.7222 & 0.7833 & 0.9389 &   -1.3000 &  -0.52220 & 0.0  \\
\hline\hline
M011 & 0.1227	&	0.0220	&	0.7151	&	0.5827	&	0.9357  &	-1.0821	&	1.0646	&	0.000345 \\
M012 & 0.1241	&	0.0224	&	0.7472	&	0.8315	&	0.8865	&	-1.2325	&	-0.7646	&	0.001204 \\
M013 & 0.1534	&	0.0232	&	0.8098	&	0.7398	&	0.8706	&	-1.2993	&	1.2236	&	0.003770\\
M014 & 0.1215	&	0.0215	&	0.8742	&	0.5894	&	1.0151	&	-0.7281	&	-0.2088	&	0.001752\\
M015 & 0.1250	&	0.0224	&	0.8881	&	0.6840	&	0.8638	&	-1.0134	&	0.0415	&	0.002789\\
M016 & 0.1499	&	0.0223	&	0.7959	&	0.6452	&	1.0219	&	-1.0139	&	0.9434	&	0.002734\\
M017 & 0.1206	&	0.0215	&	0.7332	&	0.7370	&	1.0377	&	-0.9472	&	-0.9897	&	0.000168\\
M018 & 0.1544	&	0.0217	&	0.7982	&	0.6489	&	0.9026	&	-0.7091	&	0.6409	&	0.006419\\
M019 & 0.1256	&	0.0222	&	0.8547	&	0.8251	&	1.0265	&	-0.9813	&	-0.3393	&	0.004673\\
M020 & 0.1514	&	0.0225	&	0.7561	&	0.6827	&	0.9913	&	-1.0101	&	-0.7778	&	0.009777\\
M021 &0.1472	&	0.0221	&	0.8475	&	0.6583	&	0.9613	&	-0.9111	&	-1.5470	&	0.000672\\
M022 &0.1384	&	0.0231	&	0.8328	&	0.8234	&	0.9739	&	-0.9312	&	0.5939	&	0.008239\\
M023 &0.1334	&	0.0225	&	0.7113	&	0.7352	&	0.9851	&	-0.8971	&	0.3247	&	0.003733\\
M024 &0.1508	&	0.0229	&	0.7002	&	0.7935	&	0.8685	&	-1.0322	&	1.0220	&	0.003063\\
M025 &0.1203	&	0.0230	&	0.8773	&	0.6240	&	0.9279	&	-0.8282	&	-1.5005	&	0.007024\\
M026 &0.1224	&	0.0222	&	0.7785	&	0.7377	&	0.8618	&	-0.7463	&	0.3647	&	0.002082\\
M027 &0.1229	&	0.0234	&	0.8976	&	0.8222	&	0.9698	&	-1.0853	&	0.8683	&	0.002902\\
M028 &0.1229	&	0.0231	&	0.8257	&	0.6109	&	0.9885	&	-0.9311	&	0.8693	&	0.009086\\
M029 &0.1274	&	0.0228	&	0.8999	&	0.8259	&	0.8505	&	-0.7805	&	0.5688	&	0.006588\\
M030 &0.1404	&	0.0222	&	0.8232	&	0.6852	&	0.8679	&	-0.8594	&	-0.4637	&	0.008126\\
M031 & 0.1386	&	0.0229	&	0.7693	&	0.6684	&	1.0478	&	-1.2670	&	1.2536	&	0.006502\\
M032 & 0.1369	&	0.0215	&	0.8812	&	0.8019	&	1.0005	&	-0.7282	&	-1.6927	&	0.000905\\
M033 & 0.1286	&	0.0230	&	0.7005	&	0.6752	&	1.0492	&	-0.7119	&	-0.8184	&	0.007968\\
M034 & 0.1354	&	0.0216	&	0.7018	&	0.5970	&	0.8791	&	-0.8252	&	-1.1148	&	0.003620\\
M035 & 0.1359	&	0.0228	&	0.8210	&	0.6815	&	0.9872	&	-1.1642	&	-0.1801	&	0.004440\\
M036 & 0.1390   &       0.0220  &       0.8631  &       0.6477  &       0.8985  &       -0.8632 &       0.8285  &       0.001082\\
\end{tabular}
\end{center}
\end{table*}


\begin{thebibliography}{99}

\bibitem[{{Ade et al.}(2015)}]{ade15}
Ade,~P.A.R. et al. (Planck Collaboration), 2016, Astron. \&
Astrophys. 594, A13

\bibitem[{{Agarwal \& Feldman}(2011)}]{agarwal}
Agarwal, S. \& Feldman, H., 2011,
Mon.\ Not.\ Roy.\ Astron.\ Soc.\ 410, 1647

\bibitem[{{Agarwal et al.}(2014)}]{agarwal14}
Agarwal, S., Abdalla, F.B., Feldman, H.A., Lahav, O., \& Thomas, S.A.,
2014, 
Mon.\ Not.\ Roy.\ Astron.\ Soc.\ 439, 2102

\bibitem[{{Anderson et al.}(2014)}]{anderson14}
Anderson,~L. et al., 2014, Mon.\ Not.\ Roy.\ Astron.\ Soc.\ 441, 24 

\bibitem[{{Bergner}(2011)}]{Bergner:2011}
Bergner, S., 2011, {\em Making choices in multi-dimensional parameter
  spaces}, {\em PhD thesis}, Simon Fraser University 

\bibitem[{{Bird et al.}(2012)}]{bird}
Bird, S., Viel, M., \& Haehnelt, M.G., 2012,
Mon.\ Not.\ Roy.\ Astron.\ Soc.\ 420, 2551

\bibitem[{{Brandbyge et al.}(2008)}]{brandbyge1}
Brandbyge, J., Hannestad, S., Haugbolle, T., \& Thomsen, B., 2008,
JCAP 0808, 020

\bibitem[{{Brandbyge \& Hannestad}(2009)}]{brandbyge2}
Brandbyge, J. \& Hannestad, S., 2009,
JCAP 0905, 002

\bibitem[{{Brandbyge \& Hannestad}(2010)}]{brandbyge3}
Brandbyge, J. \& Hannestad, S., 2010,
JCAP 1001, 021

\bibitem[{Banerjee \& Dalal}(2016)]{banerjee}
Banerjee, A. \& Dalal, N., 2016, JCAP 1611, 015

\bibitem[{Caldwell \& Kamionkowski}(2009)]{caldwell09}
Caldwell,~R. \& Kamionkowski, 2009, Ann.~Rev.~Nuc.~Part.~Sci. 59, 397  

\bibitem[{{Casarini et al.}(2016)}]{casarini}
Casarini, L., Bonometto, S. A., Tessarotto, E., \& Corasaniti, P.-S.,
2016, arXiv:1601.07230 

\bibitem[{{Castorina et al.}(2015)}]{castorina15}
 Castorina, E.,  Carbone, C., Bel, J., Sefusatti, E., \& Dolag, K 2015,
JCAP 1507, 043

\bibitem[{{Chevalier \& Polarski}(2001)}]{chevalier}
Chevalier, M. \& Polarski, D. 2001, Int. J. Mod. Phys. D 10, 213

\bibitem[{{Chiang et al.}(2016)}]{chiang16}
Chiang,~C.T., Li,~Y., Hu,~W., \& LoVerde,~M., 2016, Phys. Rev. D 94, 123502

\bibitem[{Dark Energy Survey Collaboration}(2016)]{DES_cosmic_shear}
  The Dark Energy Survey Collaboration et al., 2016, Phys. Rev. D, 94,
  022001 

\bibitem[{Eifler et al.}(2015)]{eifler}
  Eifler,~T., Krause,~E.,~Dodelson,~S, Zentner,~A.R., Hearin,~A.P.
  \& Gnedin,~N.Y., 2015, MNRAS, 454, 2451

\bibitem[{Feng}(2010)]{feng10}
Feng,~J.L., 2010, Ann.~Rev.~Astron.~Astrophys. 48, 495

\bibitem[{{Francis, Lewis, \& Linder}(2007)}]{francis07}
Francis, M.J., G.F. Lewis, G.F., \& Linder, E.V., 2007,
Mon.\ Not.\ Roy.\ Astron.\ Soc.\ 380, 1079 

\bibitem[{{Gardini et al.}(1999)}]{gardini}
Gardini, A., Bonometto, S.A., \& Murante, G., 1999,
ApJ 524, 510

\bibitem[{{Habib et al.}(2007)}]{HHHNW}
Habib, S., Heitmann, K., Higdon, D., Nakhleh, C., \& Williams, B., 2007,
Phys. Rev. D 76, 083503

\bibitem[{{Habib et al.}(2016)}]{habib16} 
Habib, S., Pope, A., Finkel, H., Frontiere, N., Heitmann, K., Daniel, D., Fasel, P.,
Morozov, V., Zagaris, G., Peterka, T., Vishwanath, V., Luki\'c, Z., 
Sehrish, S., \& Liao, W.-k., 2016, 
New Astronomy, 42, 49

\bibitem[{{Heitmann et al.}(2006)}]{HHHN}
Heitmann, K., Higdon, D., Nakhleh, C., \& Habib, S., 2006, ApJ 646, L1 

\bibitem[{{Heitmann et al.}(2009)}]{coyote2}
  Heitmann~K., Higdon~D., White~M., Habib~S., Williams, B.J., \&
Wagner, C.,  2009, ApJ 705, 156 

\bibitem[{{Heitmann et al.}(2010)}]{coyote1}
  Heitmann~K., White~M., Wagner~C., Habib~S., Higdon~D., 2010  
  ApJ 715, 104
 
\bibitem[{{Heitmann et al.}(2014)}]{emu_ext}
  Heitmann~K., Lawrence, E., Kwan, J., Habib, S., \& Higdon, D., 2014 
  ApJ 780, 111
  
\bibitem[{{Heitmann et al.}(2015)}]{heitmann15}
Heitmann, K., Bingham, D., Lawrence E.,  Bergner, S., et al., 2016,
ApJ 820, 108

\bibitem[{{Hu et al.}(2016)}]{hu16}
Hu,~W., Chiang,~C.T., ~Li,~Y., \& LoVerde,~M., 2016, Phys. Rev. D 94, 023002

\bibitem[{{Inman et al.}(2015)}]{inman15}
Inman, D., Emberson, J. D., Pen, U.-L., Farchi, A., Yu, H.-R., \&
Harnois-Déraps, J., 2015, 
Phys. Rev. D 92, 023502 

\bibitem[{{Joyce et al.}(2015)}]{joyce15}
Joyce,~A., Jain,~B., Khoury,~J., \& Trodden,~M., 2015, Phys. Rep.,
568, 1

\bibitem[{{Klypin et al.}(1993)}]{klypin} Klypin, A., Holtzman, J.,
  Primack, J., \& Regos, E., 1993, ApJ 416, 1

\bibitem[{Kitching et al.}(2014)]{kitching}
  Kitching,~T., Heavens,~A.~F., Alsing,~J., Erben,~T., Heymans,~C.,
  Hildebrandt,~H., Hoekstra, H., Jaffe,~A. et al., 2014, MNRAS, 442, 1326

\bibitem[{{Kwan et al.}(2013)}]{kwan13}
Kwan, J., Bhattacharya, S., Heitmann, K., \& Habib, S. 2013,
ApJ 768, 123

\bibitem[{{Kwan et al.}(2015)}]{kwan14}
  Kwan, J., Heitmann, K., Habib, S., Padmanabhan, N., Finkel, H.,
  Frontiere, N. \& Pope, A., 
  2015, ApJ 810, 35

\bibitem[{Kwan et al.}(2017)]{kwan17}
  Kwan,~J., Sanchez,~C., Clampitt,~J., Blazek,~J., Crocce,~M.,
  Jain,~B., Amara,~A., Becker~.M.~R. et al., 2017, MNRAS, 464, 4045

\bibitem[{{Lawrence et al.}(2010)}]{coyote3}
Lawrence, E., Heitmann, K., White, M., Higdon, D., Wagner, C., Habib,
S., \& Williams, B.  
2010 ApJ 713, 1322

\bibitem[{{Lewis et al.}(2000)}]{camb}
Lewis, A., Challinor, A., \& Lasenby, A. 2000, ApJ 538, 473 

\bibitem[{{Linder}(2003)}]{linder}
Linder, E. 2003, Phys. Rev. Lett. 90, 091301

\bibitem[{MacCrann et al.}(2014)]{maccrann}
  MacCrann,~N., Zuntz,~J, Bridle,~S, Jain,~B, \& Becker,~M.R., 2015,
  MNRAS, 451, 2877 

\bibitem[{{Mead et al.}(2015)}]{mead15}
Mead, A.J., Peacock, J.A., Heymans, C.,  Joudaki, S., \& Heavens, A., 2015,
Mon.\ Not.\ Roy.\ Astron.\ Soc.\ 454, 1958

\bibitem[{{Mead et al.}(2016)}]{mead16}
Mead, A.J., Heymans, C., Lombriser, L., Peacock, J.A., Steele, O.I.,
\& Winther, H.A., 2016, 
Mon.\ Not.\ Roy.\ Astron.\ Soc.\ 459, 1468

\bibitem[{{O'Shea et al.}(2010)}]{enzo}
O'Shea B.W., Bryan G., Bordner J., Norman M.L., Abel
T.,  Harkness  R.,  Kritsuk  A.,  2010,  Astrophysics  Source
Code Library, 10072

\bibitem[{{Pietroni}(2008)}]{timeRG1}
Pietroni, M. 2008, JCAP 10, 036


\bibitem[{{Schneider et al.}(2016)}]{schneider}
Schneider, A., Teyssier, R., Potter, D., Stadel, J., Onions, J., Reed,
D.S., Smith, R.E., Springel, V., Pearce, F.R., \& Scoccimarro, R., 
2016, JCAP 04, 047

\bibitem[{{Smith et al.}(2003)}]{smith03}
Smith, R. E., Peacock, J. A., Jenkins, A., et al. 2003, Mon.\ Not.\
Roy.\ Astron.\ Soc.\ 341, 1311 

\bibitem[{{Sunayama et al.}(2016)}]{sunayama16}
Sunayama, T., Padmanabhan, N., Heitmann, K., Habib, S., \& Rangel, E., 2016,
JCAP 05, 051

\bibitem[{{Takahashi et al.}(2012)}]{takahashi12}
Takahashi, R., Sato, M., Nishimichi, T., Taruya, A., \& Oguri, M., 2012,
ApJ 761, 152

\bibitem[{{Upadhye et al.}(2014)}]{upadhye14}Upadhye, A., Biswas, R., Pope, A.,
Heitmann, K., Habib, S., Finkel, H., Frontiere, N., \& Pope, A. 2014
Phys. Rev. D 89, 103515

\bibitem[{{Viel et al.}(2010)}]{viel}
Viel, M. Haehnelt, M.G., \& Springel, V., 2010,
JCAP 1006, 015

\bibitem[{{Zentner et al.}(2013)}]{zentner}
Zentner, A.R., Semboloni, E., Dodelson, S., Eifler, T., Krause, E., \& Hearin, A.
2013, Phys. Rev. D 87, 043509

\end{thebibliography}
\end{document}